\documentclass[10pt,groupedaddress]{iopart}
\usepackage{graphicx,rotating,subfigure} %% ,amsmath,amsfonts,amssymb,delarray,cite,mcite}
\usepackage{cite}
\usepackage{iopams}  
\usepackage{dsfont}

\newcommand{\im}{\mbox{i}}
\newcommand{\be}{\begin{eqnarray}}
\newcommand{\ee}{\end{eqnarray}}
\newcommand{\bulk}{{\mbox{\tiny bulk}}}

\def\nn{\nonumber}

\def\l{\left}
\def\r{\right}
\def\e{\mbox{e}}

\eqnobysec
\begin{document}
\bibliographystyle{jpa} 
\title{Thermodynamics of impurities in the
  anisotropic Heisenberg spin-1/2 chain }

\author{J. Sirker}
\address{Max-Planck Institute for Solid State Research, Heisenbergstr.~1,
  70569 Stuttgart, Germany}
\author{S. Fujimoto}
\address{Dept. of Physics,
Kyoto University, Kyoto 606-8502, Japan}
\author{N. Laflorencie}
\address{Laboratoire de Physique des Solides,
Universit\'e Paris-Sud, UMR-8502 CNRS, 91405 Orsay, France}
\address{Laboratoire de Physique Th\'eorique, IRSAMC, Universit\'e  Paul Sabatier,
CNRS, 31062 Toulouse, France}
%% \address{Dept. of Physics and Astronomy, Univ. of British Columbia, Vancouver, BC, Canada V6T 1Z1}
%% NEW ADDRESS !
\author{S. Eggert}
\address{Dept. of Physics, University of Kaiserslautern,
D-67663 Kaiserslautern, Germany}
\author{I. Affleck}
\address{Dept. of Physics and Astronomy, Univ. of British Columbia, Vancouver, BC, Canada V6T 1Z1}

%\date{\today}
\begin{abstract}
  The thermodynamics of finite open antiferromagnetic $XXZ$ chains is
  studied using field theory, Bethe Ansatz and quantum Monte Carlo
  methods. For the susceptibility %% and the specific heat
  a parameter-free result as a function of the number of sites $L$ and
  temperature $T$ beyond the scaling limit is derived. The limiting
  cases $T/J\gg 1/L$ (J being the exchange constant), where the
  boundary correction shows a logarithmically suppressed Curie
  behaviour, and $T/J\ll 1/L$, where a crossover to the ground state
  behaviour takes place, are discussed in detail.  Based on this
  analysis we present a simple formula for the averaged susceptibility
  of a spin-$1/2$ chain doped with non-magnetic impurities. We show
  that the effective Curie constant has a highly non-trivial
  temperature dependence and shows scaling in the low-temperature
  limit. Finally, corrections due to intra- and interchain couplings
  and implications for experiments on
  Sr$_2$Cu$_{1-x}$Pd$_x$O$_{3+\delta}$ and related compounds are
  discussed.
\end{abstract}

\maketitle

\section{Introduction} \label{intro}
The spin-1/2 Heisenberg chain is probably the first and most studied example
of a strongly correlated quantum system.  The first foundations towards exact
solutions date back to the early days of quantum mechanics \cite{Bethe} and
large scale numerical simulations were already performed in the early sixties
\cite{BonnerFisher}. Due to the interesting behaviour of this simple model and
the connection with many low-dimensional antiferromagnetic materials, the
spin-1/2 chain is still heavily studied today. Some of the most fundamental
properties, such as the exact amplitude of the power-law correlation functions
\cite{Affleck98,Lukyanov,LukyanovTerras} and the full temperature behaviour of the
susceptibility \cite{egg94,Lukyanov}, have only been established after the
discovery of high temperature superconductivity has lead to renewed interest
in low-dimensional antiferromagnetism. More recently, corrections from
impurities and boundaries have come into focus
\cite{EggertAffleck92,AsakawaMatsuda,WesselHaas,FujimotoEggert,FurusakiHikihara,EggertAffleckHorton,BrunelBocquet,AffleckQin,EggertRommer,
  RommerEggert,RommerEggert2,EggertAffleck95,FHLE,AsakawaSuzuki96a,AsakawaSuzuki96b,frah97,Fujimoto,BortzSirker,SirkerBortzJSTAT,SirkerLaflorencie},
which will also be the topic of this paper.

Boundary thermodynamics are especially relevant for experiments
\cite{AmiCrawford,MotoyamaEisaki,Kojima,TakigawaMotoyama,ThurberHunt} since
even the most carefully prepared samples contain imperfections which
effectively cut the chains, and corrections to the thermodynamic limit become
especially large at low temperatures. In order to analyse the pure and the
impurity contributions it is useful to define the finite size corrections
\begin{equation} 
\delta F_{FS}(L) = F(L) -  L f_\bulk \;\;\mbox{and}\;\;\delta\chi_{FS}(L) = \chi(L) -  L \chi_\bulk 
\label{deltaf} 
\end{equation}
where $F(L)$ ($\chi(L)$) is the total free energy (susceptibility) of a system
with $L$ sites and $f_\bulk$ ($\chi_\bulk$) is the free energy
(susceptibility) per site of an infinite pure system.  In the limit of $L\to
\infty$, $\delta F_{FS}$ and $\delta \chi_{FS}$ become $L$-independent
thermodynamic boundary contributions, which will be denoted by $F_B$ and
$\chi_B$ and discussed in detail later on. However, in general $L$ will also
take on small values in experimental systems according to a random
distribution.

It is well understood that finite chains with an odd number of sites
$L$ have doublet ground states, which leads to a diverging Curie
susceptibility at temperatures below the first excitation energy $ T/J
\ll 1/L$ \cite{EggertAffleck92}. %%,asa,haas}
Recently, it has been established that the boundary susceptibility
$\chi_B(T) \equiv \delta\chi_{FS}(T,L\to\infty)$ is also divergent in the
limit $T\to 0$, albeit with a Curie factor which has a logarithmic
temperature dependence \cite{FujimotoEggert,FurusakiHikihara}. For
experimental systems it is desirable to know the complete crossover
between the two limits, which has to be averaged according to a random
distribution.  If the impurity concentration is $p$ and we assume a
Poisson distribution, i.e., that the impurity positions are
uncorrelated, then the average susceptibility becomes
\cite{EggertAffleckHorton,SirkerLaflorencie}
\begin{equation} 
\label{chi_p}
\chi_p = p^2 \sum_{L=0}^\infty (1-p)^{L} \chi(L) = (1-p) \chi_\bulk + p\ \delta \bar \chi_{FS}, 
\end{equation}
where $\delta \bar \chi_{FS} = p \sum_L (1-p)^{L} \delta \chi_{FS}(L)$
and we have assumed that each impurity effectively removes one site.
The typical approximation in experimental papers to estimate the
impurity concentration by the low-temperature Curie tail corresponds
to setting $\delta \bar \chi_{FS} = 1/(8T)$ (assuming that each chain
segment with odd length acts effectively like a free
spin-$1/2$). This, however, does not capture the full complexity of
the problem and might, in particular, lead to an underestimate of the
impurity concentration by an order of magnitude.
%% and may modify the data at higher temperatures.

Several experiments have been studying the prototypical spin-$1/2$
chain compound Sr$_2$CuO$_3$
\cite{AmiCrawford,MotoyamaEisaki,TakigawaMotoyama,ThurberHunt} and the
doped system Sr$_2$Cu$_{1-x}$Pd$_x$O$_3$ \cite{Kojima} where palladium
serves as a non-magnetic impurity. The analysis of the susceptibility
even in the undoped system has been hampered by large Curie-like
contributions at low temperatures making a detailed test of the
theoretical predictions for the susceptibility of the pure system
\cite{egg94} difficult. In \cite{AmiCrawford,MotoyamaEisaki} it has
been shown that these contributions can be largely suppressed by
annealing the samples. It therefore has been suggested that the
Curie-like contribution is caused by excess oxygen. Each excess oxygen
atom would then dope two holes into the chain. A detailed theoretical
analysis of the susceptibility data \cite{SirkerLaflorencie} has shown
that it seems to be possible to consistently describe the data by
assuming that these holes are basically immobile and therefore
effectively cut the chain into finite segments. A more detailed
understanding of these experiments is desirable because it might also
shed some light on the general issue of oxygen doping in cuprates, and
in particular, in the high-temperature superconductors.

% However, at least for the powder samples studied by Ami {\it et al.}  an
% alternative explanation was proposed based on the observation that
% Sr$_2$CuO$_3$ decomposes into Sr$_2$Cu(OH)$_6$ under exposure to air and water
% \cite{HillJohnston}. In any case, suppressing Curie-like contributions for the
% pure system as far as possible by annealing or other techniques seems to be
% essential to test the theoretical predictions for the susceptibility of the pure
% Heisenberg chain \cite{egg94} as well as the theoretical predictions for the
% averaged susceptibility of the doped system as presented in this work in more
% detail.

Our paper is organised as follows: In section \ref{fieldtheory} the
effective field theory description of the spin-1/2 chain is
reviewed. In section \ref{general} we calculate the free energy for an
open anisotropic Heisenberg chain of finite length at finite
temperature by field theory methods beyond the scaling limit. Next, we
consider the limit $L\to\infty$ with $T/J$ fixed in section
\ref{boundary} where $\delta F_{FS}$ and $\delta \chi_{FS}$ as defined
in (\ref{deltaf}) become well defined boundary quantities. In section
\ref{IsoCase} we then discuss separately the isotropic model which is
the most interesting case from an experimental point of view. In the
limit, $T/J\ll 1/L$, considered in section \ref{ground_state}, a
crossover to the ground state limit takes place and logarithmic
corrections to the scaling limit result are studied in detail. In
section \ref{avg_susci} we then derive a simple formula which can be
used to fit the experimentally measured averaged susceptibility
$\chi_p$. We also discuss here the temperature dependence of the
effective Curie constant $C\sim T\delta\bar{\chi}_{FS}$. In any real
material there will be residual couplings bridging between the chain
segments (intrachain couplings) as well as interchain couplings which
will usually lead to some sort of magnetic order at low
temperatures. The effect of such couplings will be discussed in
section \ref{seg-couplings}. A comparison between our theoretical
predictions and experiments on Sr$_2$Cu$_{1-x}$Pd$_x$O$_3$
\cite{Kojima} is presented in section \ref{experiment}. Finally, we
conclude and give a short summary in section \ref{conclusions}.

%% Section \ref{crossover} discusses the behaviour for finite
%% chains by combining field theory results with numerical data.  In section
%% \ref{conclusion} we conclude and discuss the implications for experiments.  A
%% explicit procedure for fitting the impurity contribution is presented which
%% can be used to determine the concentration $p$.

\section{Effective field theory for the spin-1/2 chain} 
\label{fieldtheory}
The Hamiltonian for the spin-1/2 Heisenberg chain with open boundaries in an
external magnetic field $h$ is given by,
\begin{equation}
H=J\sum_{i=1}^{L-1}
[S_{i}^xS_{i+1}^x+S_{i}^yS_{i+1}^y+\Delta S_{i}^zS_{i+1}^z]-h\sum_{i=1}^L
S_{i}^z,
\label{xxz}
\end{equation}
with $J>0$ for an antiferromagnetic system. Here, we have introduced an
exchange anisotropy $\Delta$ which is convenient to obtain a low-energy
effective theory by bosonisation.  Such a description is by now well
established \cite{Affleck_lesHouches,Lukyanov} and numerical simulations have
shown its applicability for $L \gtrsim 10$ and low temperatures $T \lesssim
0.1J$, which is also the most interesting regime for experiments. In the
massless case, $-1<\Delta \leq 1$, the low energy fixed point of (\ref{xxz})
is the Tomonaga-Luttinger liquid, which belongs to the universality class of
the Gaussian theory with the central charge $c=1$.

Ignoring irrelevant operators the Hamiltonian is then equivalent to a free boson model
\begin{equation}
\label{H0}
H_0 = \frac{v}{2}\int_0^{La'}\!\! dx \; [\Pi^2 +(\partial_x\phi)^2]
-h\sqrt{\frac{K}{2\pi}}\int^{La'}_0 \!\! dx \;  \partial_x\phi, 
\end{equation}
where $a'$ is the lattice constant. $\phi$ is a bosonic field and
$\Pi=v^{-1}\partial_t\phi$ the conjugate momentum obeying the standard
commutation rule $[\phi(x),\Pi(x')]=\im\delta(x-x')$. The dependence of the
Luttinger parameter $K$ on anisotropy $\Delta$ is known exactly from Bethe
Ansatz
\begin{equation}
K=[1-\frac{1}{\pi}\cos^{-1}(\Delta)]^{-1}, 
\label{lpa}
\end{equation}
as well as the velocity \cite{tak99}
\begin{equation}
v =\frac{JK\sin(\pi/K)}{2(K-1)}a'\; .
\end{equation}
In the following we will set the lattice constant $a'\equiv 1$. Factors of $a'$
can be easily restored in the final results by dimensional analysis. Using the
mode expansion for open boundary conditions (OBCs) 
\begin{eqnarray}
\label{ModeExp}
\fl \phi(x,t) &=& \sqrt{\frac{\pi}{8K}} +\sqrt{\frac{2\pi}{K}} S_z \frac{x}{L} + \sum_{n=1}^\infty
\frac{\sin\l(\pi nx/L\r)}{\sqrt{\pi n}}\l(\e^{-i\pi n \frac{vt}{L}}a_n +
\e^{i\pi n\frac{vt}{L}}a_n^\dagger\r) 
%% \e^{i\pi n\frac{vt}{L}}a_n^\dagger   
\end{eqnarray}
the Hamiltonian can also be expressed in terms of bosonic creation and
annihilation operators $a_n$
\begin{equation}
H_0 = \frac{\pi v}{L K}S_z^2 + \frac{\pi v}{L}\sum_{n=1}^\infty n (a_n^\dagger a_n^{~}+1/2) - h S_z.
\end{equation}
The zero mode operator $S_z = \sqrt{\frac{K}{2\pi}}\int^L_0  dx\ \partial_x\phi$ measures 
the 
total magnetisation of the chain and is therefore quantised to take on integer values 
for chains with an even number of sites $L$ and half-integer values for odd $L$.

Up to this point the effective model is easily solvable exactly.  The partition function 
can be obtained directly by summing over all eigenvalues
\begin{eqnarray}
\label{part}
Z_0 &=& \sum_{S_z}e^{-\frac{\pi v}{KLT}S_z^2+ \frac{h}{T}S_z}
\prod_{n=1}^\infty \l[2\sinh\l(\frac{\pi v}{2LT}n\r)\r]^{-1} \\
&=& \theta\left(e^{-\frac{\pi v}{K LT}}, -i\frac{h}{2T}\right) \prod_{n=1}^\infty \l[2\sinh\l(\frac{\pi v}{2LT}n\r)\r]^{-1} \nonumber
\end{eqnarray}
where $\theta(q,u)$ is the elliptic theta function of the third kind $\theta =
\theta_3(q,u) = \sum_{n=-\infty}^{\infty} q^{n^2} e^{i2 n u}$ for integer
$S_z$ (even $L$) and of the second kind $\theta = \theta_2(q,u) =
\sum_{n=-\infty}^{\infty} q^{(n+1/2)^2} e^{i(2 n+1) u}$ for half-integer $S_z$
(odd $L$).

The free energy per site is then given by
\begin{equation}
\label{freeE}
f_0 = - \frac{T}{L} \l\{\ln \theta\left(e^{-\frac{\pi v}{K LT}},
  -i\frac{h}{2T}\right) - \sum_{n=1}^\infty \ln \l[2\sinh\l(\frac{\pi v}{2LT}n\r)\r]\r\} \; .
\end{equation}
and the susceptibility by 
\begin{eqnarray} 
\label{scalechi}
\fl \chi_0 &=& -\frac{\partial^2 }{\partial h^2}\bigg|_{h=0}f_0 = \frac{1}{LT} \frac{\sum_{S_z} S_z^2 
e^{-\frac{\pi v}{KLT}S_z^2}}{\sum_{S_z} e^{-\frac{\pi v}{KLT}S_z^2}} 
= -\frac{1}{4LT}\frac{\partial^2}{\partial u^2}\bigg|_{u=0} \ln
\theta\left(e^{-\frac{\pi v}{KLT}},u\right)
\end{eqnarray}
Note, that $\chi_0$ has a simple scaling form as a function of $LT$
\cite{EggertAffleck92}.
%%  as depicted in Fig.~\ref{scalechifig} for the isotropic
%% case.
% \begin{figure} 
% \begin{center}
% \includegraphics*[width=0.99\columnwidth]{scalingpart.eps}
% \caption{The scaling behaviour of the susceptibility as a function of
%   $LT/v$ in Eq.~(\ref{scalechi}) for the isotropic case $\Delta=1, \
%   K=1, \ v=\pi/2$ (black solid lines). The red dashed lines represent
%   the asymptotics given in Eq.~(\ref{limits}).  At low temperatures
%   the susceptibility is exponentially suppressed for even $L$ and
%   Curie divergent for odd $L$.}
% \label{scalechifig}
% \end{center}
% \end{figure}
The following limiting cases are of interest
\begin{eqnarray}
\label{limits}
\chi_0 &=& \l\{ \begin{array}{ll}
\frac{2}{LT}\exp\l[-\frac{\pi v}{KLT}\r] & LT/v\to 0,\; \mbox{$L$ even} \\[0.2cm]
\frac{1}{4TL} & LT/v\to 0,\; \mbox{$L$ odd} \\[0.2cm]
\frac{K}{2\pi v}\pm \frac{2K^2}{v}\frac{LT}{v}\exp\l(-\pi K\frac{LT}{v}\r) & LT/v\to\infty \; \; .
\end{array} \r.
\end{eqnarray}
Here the plus (minus) sign in front of the second term in the last
line corresponds to L odd (even), respectively. This term represents
the leading finite size correction to the thermodynamic limit result
$\chi_0=K/(2\pi v)$. 
% This means that at up to this order finite size
% corrections will effectively cancel out when averaging over even and
% odd chain segments if $T/v \gg 1/L$. 
In the limit $T/v \ll 1/L$ on the other hand, the susceptibility for a
chain of fixed even length vanishes exponentially because it locks
into its singlet ground state, whereas $\chi_0$ for a chain of odd
length exhibits a $1/T$ divergence due to its doublet ground state.

\begin{figure} 
\begin{center}
\includegraphics*[width=0.99\columnwidth]{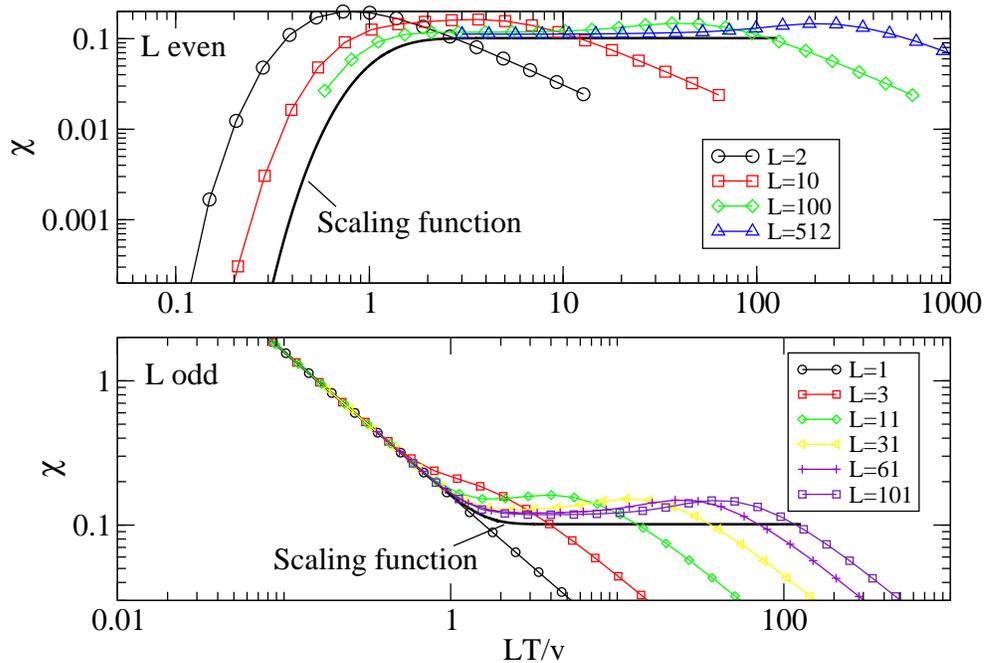}
\caption{The scaling behaviour of the susceptibility as a function of $LT/v$
  in Eq.~(\ref{scalechi}) for the isotropic case $\Delta=1, \ K=1, \ v=\pi/2$
  (black solid lines) for $L$ even (upper panel) and $L$ odd (lower panel),
  respectively. The symbols denote quantum Monte Carlo data for chains of
  different lengths, obtained within the SSE loop algorithm framework~
  \cite{Sandvik2002}.}
\label{scalechifig}
\end{center}
\end{figure}

If we were only interested in the qualitative behaviour we could stop
at Eq.~(\ref{scalechi}), which in some cases is sufficient for
agreement with experimental data \cite{AsakawaMatsuda}. However, the
comparison with quantum Monte Carlo data in Fig.~\ref{scalechifig}
clearly shows sizeable deviations from the scaling limit result.
For more accurate predictions and the understanding of impurity
corrections it is therefore necessary to take the leading irrelevant
operator into account, which is given by
\cite{Affleck_lesHouches,Lukyanov}
\begin{equation}
\delta H=\lambda\int_0^L dx \,\cos(\sqrt{8\pi K}\phi).
\label{int1}
\end{equation}
The operator $\cos(\sqrt{8\pi K}\phi)$ has scaling dimension $2K$ so
that this perturbation becomes marginally irrelevant at the isotropic
point $K=1$. The next-leading irrelevant operators at zero magnetic
field are of the form $\sim (\partial_x\phi)^4$ and will be neglected
in the following unless it is explicitly stated otherwise. The
coupling strength $\lambda$ has been obtained exactly by Bethe ansatz
\cite{Lukyanov} and is given in a particular normalisation as
discussed in detail later on by
\begin{equation}
\label{B108}
\lambda
=\frac{K\Gamma(K)\sin\pi/K}{\pi\Gamma(2-K)}\l[\frac{\Gamma\l(1+\frac{1}{2K-2}\r)}{2\sqrt{\pi}\Gamma\l(1+\frac{K}{2K-2}\r)}\r]^{2K-2}
\; .
\end{equation} 
At the isotropic point $\Delta=1$ ($K=1$) where the interaction (\ref{int1})
becomes marginal the amplitude $\lambda$ as given by (\ref{B108}) vanishes.
Here $\lambda$ has to be replaced by an RG-improved coupling constant, and, in
general, becomes a function of the different scales $T, L$, and $h$. This will
be discussed in detail in section \ref{IsoCase}.

%% , the effective low energy theory is
%% described by the level $k=1$ SU(2) Wess-Zumino-Witten model 
%% with a marginally irrelevant interaction
%% \begin{eqnarray}
%% H&=&H_{WZW}+H_m, \label{eff2}\\
%% H_m&=&-g\int_0^L\frac{dx}{2\pi}\sum_{a=1}^3J^a(x)\bar{J}^a(x).\label{int2}
%% \end{eqnarray}
%% Here $H_{WZW}$ is the Hamiltonian of the level $k=1$ SU(2)
%% Wess-Zumino-Witten model, and $J^a(x)$ ($\bar{J}^a(x)$)
%% is the left (right) moving current of the level $k=1$ SU(2) 
%% Kac-Moody algebra.
%% The running coupling constant $g$ depends on
%% temperature $T$ and an external magnetic field $h$ through the scaling 
%% equation,\cite{luk}
%% \begin{eqnarray}
%% g^{-1}+\frac{1}{2}\ln(g)=-{\rm Re}[\psi(1+\frac{i h}{2\pi T})]
%% +\ln(\sqrt{\pi/2} e^{1/4}J/T), \label{sca}
%% \end{eqnarray}
%% with $\psi(x)$ the di-gamma function.

Generally, in systems with boundaries, there may be boundary operators
in addition to bulk interactions.  However, as was pointed out in
Ref.~\cite{BrunelBocquet}, in the absence of symmetry-breaking
external fields at boundaries, there are no relevant boundary
operators for the anisotropic Heisenberg chain.  The leading
irrelevant boundary operator has scaling dimension $2$ \footnote{Note
  that the critical dimension for a boundary operator in $(1+1)$
  dimensions is $1$, whereas it is $2$ for bulk operators.} and is of
the form $\delta(x)(\partial_x \phi)^2$ \cite{EggertAffleck92}. This
operator effectively leads to a replacement of the length $L$ of the
system by some effective length $L'$ as will be discussed in the next
section.

\section{Thermodynamics for a finite chain beyond the scaling limit}
\label{general}
To calculate the free energy and susceptibility for an open $XXZ$ chain with
length $L$ at finite temperature $T$ beyond the scaling limit
(\ref{freeE},\ref{scalechi}), the leading correction (\ref{int1}) has to be
taken into account. As this operator is irrelevant for $0\leq\Delta < 1$ it is
sufficient to use perturbation theory. This anisotropic regime will be
considered here. 
% Experimentally most interesting is the isotropic case $\Delta
% = 1$ where this correction becomes marginal. We will discuss this case
% separately in section \ref{IsoCase}.

In first order perturbation theory we obtain
\begin{equation}
\label{delta_f}
\delta f_1 =\frac{\lambda}{L} \int_0^L dx
\l\langle\cos\l(\sqrt{8\pi K}\phi\r)\r\rangle \; . 
\end{equation}
This term has been considered in
Refs.~\cite{FujimotoEggert,FurusakiHikihara,SirkerBortzJSTAT} in the
limit $L\to\infty$ and gives the leading contribution to the boundary
free energy. A bulk contribution, i.e.~a term which scales linearly
with $L$ at large $L$, occurs in second order perturbation theory in
$\lambda$ \cite{Lukyanov} (see also \ref{app_A}).

The expectation value can be split into an $S_z$-part (zero mode) and an oscillator part
\begin{eqnarray}
\label{expvalue}
\l\langle \exp\l(\pm\im\sqrt{8\pi K}\phi\r)\r\rangle &=& \l\langle
\exp\l(\pm\im\sqrt{8\pi K}\phi\r)\r\rangle_{\mbox{\tiny $S_z$}} \l\langle
\exp\l(\pm\im\sqrt{8\pi K}\phi\r)\r\rangle_{\mbox{\tiny osc.}} \nonumber \\
& \!\!\!\!\!\!\!\!\!\!\!\!\!\!\!\!\!\!\!\!\!\!\!\!\!\!\! =&
 \!\!\!\!\!\!\!\!\!\!\!\!\!\! \l\langle\exp\l(\pm\im\sqrt{8\pi K}\phi\r)\r\rangle_{\mbox{\tiny $S_z$}} 
\exp\l(-4\pi K\langle\phi\phi\rangle_{\mbox{\tiny osc.}} \r) \, ,
\end{eqnarray}
where we used the cumulant theorem for bosonic modes in the second line. Using
the mode expansion (\ref{ModeExp}) we find
\begin{equation}
\label{Sz-part}
\l\langle \exp\l(\pm\im\sqrt{8\pi K}\phi\r)\r\rangle_{\mbox{\tiny $S_z$}} =
  \frac{\sum_{S_z}\e^{\pm \frac{4\pi i S_z x}{L}}\e^{-\frac{\pi
  vS_z^2}{KLT}}\e^{\frac{hS_z}{T}}}{\sum_{S_z}\e^{-\frac{\pi vS_z^2}{KLT}}\e^{\frac{hS_z}{T}}} 
\end{equation}
and
\begin{equation}
\label{osc1}
\langle\phi\phi\rangle_{\mbox{\tiny osc.}} = \sum_{l=1}^\infty
\frac{\sin^2(\pi lx/L)}{\pi l}\l(1+\frac{2}{\e^{\pi vl/(TL)}-1}\r) \; .
\end{equation}
The zero temperature part of (\ref{osc1}) is divergent and we have to
introduce a cutoff $\alpha$ with dimensions of length, of order a
lattice spacing. Doing so and using $\sum_{l=1}^{\infty} z^l/l =
-\ln(1-z)$ for $|z|<1$ leads to
\begin{eqnarray}
\label{osc2}
&&\sum_{l=1}^\infty \frac{\sin^2(\pi lx/L)}{\pi l} \e^{-\alpha\pi l/L} =
-\frac{1}{2\pi}\ln(\alpha\pi/L) \\
&+& \frac{1}{4\pi}\ln\l[\l(1-\e^{2\pi i
  x/L}\e^{-\alpha\pi/L}\r)\l(1-\e^{-2\pi i x/L}\e^{-\alpha\pi/L}\r)\r] \; .\nonumber
\end{eqnarray}
Writing the exponential factor of the finite temperature part in
Eq.~(\ref{osc1}) as a geometric series we find
\begin{eqnarray}
\label{osc3}
&& \frac{1}{\pi}\sum_{l=1}^\infty\sum_{n=1}^\infty
\l(\frac{1}{l}-\frac{\e^{2\pi i xl/L}+\e^{-2\pi i xl/L}}{2l}\r)\e^{-\pi
  vnl/(TL)} \nonumber \\
&=&\frac{1}{\pi}\sum_{n=1}^\infty\l\{-\ln\l(1-\e^{-\pi v n/(TL)}\r)
\r. \\
&+& \l.\frac{1}{2}\ln\l[\l(1-\e^{2\pi i
  x/L}\e^{-\pi vn/(TL)}\r)\l(1-\e^{-2\pi i x/L}\e^{-\pi vn/(TL)}\r)\r] \r\}
\; .\nonumber
\end{eqnarray}
Combining Eqs.~(\ref{osc1}, \ref{osc2}, \ref{osc3}) gives
\begin{eqnarray}
\label{osc4}
&&\exp\l\{-4\pi K\langle\phi\phi\rangle_{\mbox{\tiny osc.}} \r\} \\ 
&=& \l(\frac{\alpha\pi}{L}\r)^{2K}\l[\l(1-\e^{2\pi i
  x/L}\e^{-\alpha\pi/L}\r)\l(1-\e^{-2\pi i x/L}\e^{-\alpha\pi/L}\r)\r]^{-K}
  \nonumber \\
&\times& \prod_{n=1}^\infty\l[\frac{\l(1-\e^{2\pi i
  x/L}\e^{-\pi vn/(TL)}\r)\l(1-\e^{-2\pi i x/L}\e^{-\pi
  vn/(TL)}\r)}{\l(1-\e^{-\pi v n/(TL)}\r)^2}\r]^{-2K}  \nonumber \\
&=& \l(\frac{2\alpha\pi}{L}\r)^{2K}\l[\l(1-\e^{2\pi i
  x/L}\e^{-\alpha\pi/L}\r)\l(1-\e^{-2\pi i x/L}\e^{-\alpha\pi/L}\r)\r]^{-K}
  \nonumber \\
&\times& \sin^{2K}\l(\frac{\pi x}{L}\r)\frac{\eta^{6K}\l(\e^{-\pi
    v/(TL)}\r)}{\theta_1^{2K}\l(\frac{\pi x}{L},\e^{-\pi v/(2TL)}\r)} \nonumber
\end{eqnarray}
In the last step we have written the oscillator part in terms of the Dedekind
eta-function
\begin{equation}
\label{osc5}
\eta(w) = w^{1/24} \prod_{n=1}^\infty \l(1-w^n\r)
\end{equation}
and the elliptic theta-function of the first kind
\begin{equation}
\label{osc6}
\theta_1(u,q) = 2 q^{1/4}\sin u \prod_{n=1}^\infty  \l(1-2q^{2n}\cos 2u
+q^{4n}\r)\l(1-q^{2n}\r) \; .
\end{equation}
From (\ref{Sz-part}) and (\ref{osc4}) we now directly obtain the correction to
the scaling form of the free energy (\ref{freeE})
\begin{eqnarray}
\label{freeEcorr}
\delta f_1 &=&
2\tilde{\lambda}\l(\frac{\pi}{L}\r)^{2K}\!\!\eta^{6K}\l(\e^{-\frac{\pi
    v}{TL}}\r)  \int_0^{1/2} \!\! dy \,\frac{\tilde{g}\l(y,\e^{-\frac{\pi
      v}{KLT}},e^{\frac{h}{T}}\r)}{\theta_1^{2K}\l(\pi y,\e^{-\frac{\pi v}{2TL}}\r)} \nonumber
\\
&\times& 2^{2K}\sin^{2K}\l(\pi y\r)\l[\l(1-\e^{2\pi i
 y}\e^{-\alpha\pi/L}\r)\times \mbox{h.c.}\r]^{-K}  %%\l(1-\e^{-2\pi i y}\e^{-\alpha\pi/L}\r)\r]^{-K}
\end{eqnarray}
with 
\begin{eqnarray}
\label{freeEcorr2}
\tilde{g}\l(y,q,w\r) &=& \frac{\sum_{S_z} \cos(4\pi S_z y) q^{S_z^2}w^{S_z}}{\sum_{S_z}
  q^{S_z^2}w^{S_z}} 
\end{eqnarray}
and $\tilde{\lambda} =\alpha^{2K}\lambda$. It is actually
$\tilde{\lambda}$ which is given by Eq.~(\ref{B108}), but we will
identify both in the following. Note, that the integral in
(\ref{freeEcorr}) is convergent for $K<1/2$. In this case the cutoff
$\alpha$ can be dropped in the last line in (\ref{freeEcorr}) which
then becomes equal to one. For $K>1/2$ we can isolate the cutoff
independent part by subtracting the Taylor expand of the integrand up
to sufficient order in $y$, and then take $\alpha\to 0$.

We will discuss this in detail in the following for the correction to the
susceptibility which is readily obtained from (\ref{freeEcorr}) and reads
\begin{eqnarray}
\label{suscicorr}
\delta\chi_1 &=&
\frac{2\lambda}{T^2}\l(\frac{\pi}{L}\r)^{2K}\!\!\eta^{6K}\l(\e^{-\frac{\pi
    v}{TL}}\r)  \int_0^{1/2} \!\!\!\!\!\! dy \,\frac{g_0\l(y,\e^{-\frac{\pi
      v}{KLT}}\r)}{\theta_1^{2K}\l(\pi y,\e^{-\frac{\pi v}{2TL}}\r)} \nonumber
\\
&\times& 2^{2K}\sin^{2K}\l(\pi y\r)\l[\l(1-\e^{2\pi i
 y}\e^{-\alpha\pi/L}\r)\times \mbox{h.c.}\r]^{-K}  %%\l(1-\e^{-2\pi\im y}\e^{-\alpha\pi/L}\r)\r]^{-K}
\end{eqnarray}
where we have defined a new function for the zero mode part
\begin{eqnarray}
\label{suscicorr2}
\fl g_0\l(y,q\r) &=& -\frac{\sum_{S_z} S_z^2\cos(4\pi S_z y) q^{S_z^2}}{\sum_{S_z}
  q^{S_z^2}} 
+ \frac{\l(\sum_{S_z} \cos(4\pi S_z y) q^{S_z^2}\r)\l(\sum_{S_z}
  S_z^2  q^{S_z^2}\r)}{\l(\sum_{S_z}
  q^{S_z^2}\r)^2} \, . 
\end{eqnarray} 
Because of the modified zero mode part, the integral is now convergent for
$K<3/2$ and the last line can be set equal to one in this case again. For
$3/2<K<5/2$ we can obtain a convergent integral, i.e. the cutoff independent
part, by subtracting just the first non-vanishing order in a Taylor expand in
$y$ and setting $\alpha \to 0$ then. Noting that 
\begin{equation}
\label{taylor}
\lim_{y\to 0} \frac{2^{2K}\sin^{2K}\l(\pi y\r)\eta^{6K}\l(\e^{-\frac{\pi
      v}{TL}}\r)}{\theta_1^{2K}\l(\pi y,\e^{-\frac{\pi v}{2TL}}\r)} = 1
\end{equation}
the cutoff independent part of (\ref{suscicorr}) for $3/2<K<5/2$ is given by
\begin{eqnarray}
\label{suscicorr_ind}
\!\!\!\!\!\!\!\!\!\!\!\!\!\!\!\!\!\!\!\!\delta\chi_1 &=&
\frac{2\lambda}{T^2}\l(\frac{\pi}{L}\r)^{2K}\!\!\!\! \int_0^{1/2} \!\!\!\!\!\!
dy \l\{ \frac{\eta^{6K}\l(\e^{-\frac{\pi v}{TL}}\r) g_0\l(y,\e^{-\frac{\pi
      v}{KLT}}\r)}{\theta_1^{2K}\l(\pi y,\e^{-\frac{\pi v}{2TL}}\r)}
\r. - \l. \frac{2 g_2\l(\e^{-\frac{\pi v}{KLT}}\r)}{(2\pi y)^{2K-2}} \r\} 
\end{eqnarray}
where 
\begin{equation}
\label{g2}
g_2\l(q\r) = \frac{\sum_{S_z} S_z^4  q^{S_z^2}}{\sum_{S_z}
  q^{S_z^2}} - \frac{\l(\sum_{S_z} S_z^2  q^{S_z^2}\r)^2}{\l(\sum_{S_z}
  q^{S_z^2}\r)^2} \, . 
\end{equation} 
Now we have to add again the first non-vanishing order in the Taylor expand in
$y$ but this time we keep the cutoff $\alpha$ giving us the non-universal
contribution
\begin{equation}
\label{suscicorr_nonuni}
\delta\chi_1^{nu} = \frac{16\pi^2\lambda}{T^2}\l(\frac{\pi}{L}\r)^{2K}
\!\!\!\! g_2\l(\e^{-\frac{\pi v}{KLT}}\r) \int_0^{1/2} \!\!\!\!\!\! 
\frac{y^2 \, dy}{\l[(2\pi y)^2+\l(\frac{\alpha\pi}{L}\r)^2\r]^K} \; .
\end{equation}
Shifting the upper boundary of integration to infinity, we can evaluate the
integral and find
\begin{equation}
\label{suscicorr_nonuni2}
\delta\chi_1^{nu} = \frac{\pi^{3/2}\lambda}{2T^2L^3}g_2\l(\e^{-\frac{\pi v}{KLT}}\r)\frac{\Gamma(K-3/2)}{\Gamma(K)}\alpha^{3-2K} \; .
\end{equation}
The other non-universal correction stems from the irrelevant boundary operator
$\sim\delta(x)(\partial_x\phi)^2$. Including this term into the Hamiltonian
(\ref{H0}) is equivalent to replacing the length by some effective length
$L\to L' \equiv L+a$. Using this effective length in the exponentials in
(\ref{scalechi}) and expanding to lowest order in $a$ yields
\begin{equation}
\label{NonU}
\delta \chi_2 = \frac{\pi va}{KT^2L^3} g_2\l(\e^{-\frac{\pi
  v}{KLT}}\r) \; .
\end{equation}
We see that this correction has the same form as (\ref{suscicorr_nonuni2}). We
therefore can consider $a$ in (\ref{NonU}) as an effective parameter
incorporating both corrections. In the
thermodynamic limit $g_2\l(\e^{-\pi v/(KLT)}\r) \to K^2T^2L^2/(2\pi^2v^2)$ and
$\delta \chi_2 \to Ka/(2\pi v L)$. In this limit we can compare the field
theory result with a recent calculation of the boundary susceptibility based
on the Bethe ansatz \cite{BortzSirker,SirkerBortzJSTAT} leading to
\begin{equation}
\label{NonU3}
a=2^{-1/2}\sin\l[\pi K/(4K-4)\r]/\cos\l[\pi/(4K-4)\r] \; .
%% a=\frac{1}{\sqrt{2}}\frac{\sin\frac{\pi K}{4K-4}}{\cos\frac{\pi}{4K-4}} \; .
\end{equation}
Eq.~(\ref{suscicorr}) for $K<3/2$ or (\ref{suscicorr_ind}) for $K>3/2$ taken
together with (\ref{NonU}, \ref{NonU3}) is therefore a parameter-free result
for the susceptibility of an open chain with length $L$ at temperature $T$ to
first order in the Umklapp scattering. In \cite{SirkerLaflorencie} it has been
shown by comparing with quantum Monte-Carlo data that this formula does
describe the susceptibility of open chains for $L\gtrsim 10$ and $T/J\lesssim
0.1$ very well. Note, however, that the parameter $a$ in (\ref{NonU3}) has
poles at $K=(4n+3)/(4n+2)$, $n=0,1,2,\cdots$ with an accumulation point at
$K=1$. At these special points a contribution stemming from a different
irrelevant operator will also be divergent while having the same dependence on
temperature and length and both terms taken together will give something
finite. In the limit $L\to\infty$ this has been investigated in detail in
Ref.~\cite{SirkerBortzJSTAT}. In particular, it has been shown in this limit
that at $K=3/2$ ($\Delta = 1/2$) the two contributions (\ref{suscicorr}) and
(\ref{NonU}) ``conspire'' to produce a logarithmic temperature dependence of
the boundary susceptibility.

\section{Boundary contributions: The limit $L\to\infty$ with $T/J$ fixed}
\label{boundary}
\begin{figure} 
\begin{center}
\includegraphics*[width=0.99\columnwidth]{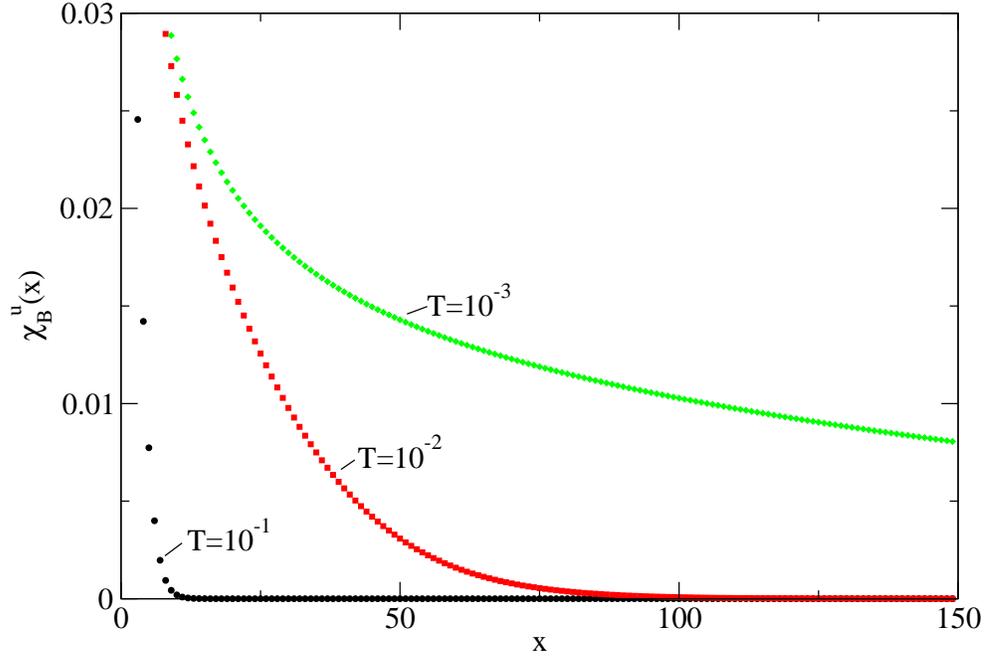}
\caption{The uniform part of the local boundary susceptibility $\chi_B^u(x)$
  following from Eq.~(\ref{delta_f_thermo}) for $K=1.2$ at different
  temperatures. The boundary susceptibility $\chi_B$ is obtained by
  integrating over $\chi_B^u(x)$ leading to a $1/T^{3-2K}$ divergence if
  $K<3/2$.}
\label{Fig_chiB_local}
\end{center}
\end{figure}

The boundary free energy can be obtained directly by starting with the
integral (\ref{delta_f}) in the limit $L\to\infty$ and using either boundary
conformal field theory \cite{FujimotoEggert} or the thermodynamic limit result
for the correlation function \cite{FurusakiHikihara,SirkerBortzJSTAT}. Here we
want to show that our more general result (\ref{freeEcorr}) for a finite chain
reduces to the known result in the thermodynamic limit. First, we consider the
limit $L\to\infty$ for the function $\tilde{g}$ in (\ref{freeEcorr2}). For
$x\ll L$ we can replace the sums by integrals and find
\begin{eqnarray}
\label{limit1}
\tilde{g}\l(\frac{x}{L},\e^{-\frac{\pi v}{KLT}},e^{\frac{h}{T}}\r) &\approx& \e^{-\frac{4\pi
    KT}{vL}x^2}\cos\l(\frac{2Khx}{v}\r) \stackrel{L\to\infty}{\to}  \cos\l(\frac{2Khx}{v}\r)
\end{eqnarray}
We see that $\tilde{g}$ decays fast away from the boundary. If
$x\sim L$, on the other hand, we can set $x\to L-x$ and because $S_z$ is
integer or half-integer $\cos(4\pi S_z (L-x)/L) = \cos(4\pi S_z x/L)$ and we
get exactly the same result again. In the thermodynamic limit the two
boundaries therefore become independent and each of them yields the same
contribution.  To perform the thermodynamic limit for the oscillator part it
is easiest to go back to Eqns.~(\ref{osc1}, \ref{osc2}, \ref{osc3}). We want
to consider here only the case $K<3/2$. In this case the cutoff in the second
line of (\ref{osc2}) can be dropped leading to
\begin{eqnarray}
\label{limit2}
\sum_{l=1}^\infty \frac{\sin^2(\pi lx/L)}{\pi l} \e^{-\alpha\pi l/L} &\approx & \frac{1}{2\pi}\ln\l\{\frac{L\l(2-2\cos\l(\frac{2\pi
    x}{L}\r)\r)^{1/2}}{\alpha \pi}\r\} \nonumber \\
&\stackrel{L\to\infty}{\to} & \frac{1}{2\pi}\ln\l\{ \frac{2x}{\alpha}\r\} \; .
\end{eqnarray}
Note that $\alpha$ will later be absorbed again into the coupling
constant $\lambda\to\tilde{\lambda}=\alpha^{2K}\lambda$. For the finite
temperature part we can replace the momentum sum in (\ref{osc3}) by an
integral, allowing us to perform the sum over $n$ (``Matsubara sum'') exactly
 \begin{eqnarray}
\label{limit3}
&& \frac{1}{\pi}\sum_{n=1}^\infty\sum_{l=1}^\infty
\l(\frac{1}{l}-\frac{\e^{2\pi i xl/L}+\e^{-2\pi i xl/L}}{2l}\r)\e^{-\pi
  vnl/(TL)} \nonumber \\
&\to&\frac{1}{\pi}\sum_{n=1}^\infty\int_0^\infty dk\,\l(\frac{1}{k}-\frac{\e^{2i
    kx}+\e^{-2i kx}}{2k}\r)\e^{-\frac{vk}{T}n} \nonumber \\
&=& \frac{1}{2\pi}\sum_{n=1}^\infty\ln\l(1+\frac{4T^2x^2}{n^2v^2}\r) =\frac{1}{2\pi}\ln\l(\frac{v}{2\pi T x}\sinh\l(\frac{2\pi Tx}{v}\r)\r) \; .
\end{eqnarray}
Thus we find in the thermodynamic limit
\begin{equation}
\label{delta_f_thermo}
F_B = L\,\delta f_1(L\to\infty) = 2\lambda\int_0^\infty \!\!\!\! dx \; \frac{\cos\l(\frac{2Khx}{v}\r)}{\l[\frac{v}{\pi
    T}\sinh\l(\frac{2\pi Tx}{v}\r)\r]^{2K}} \; .
\end{equation}
%% The amplitude $\lambda_S$ in Sebastian's draft is related to our $\lambda$
%% by $\lambda_S = (2\pi/v)\lambda$. Then all results agree.
This is the result obtained previously by boundary conformal field theory
\cite{FujimotoEggert} and by using the result for the correlation function in
the thermodynamic limit in (\ref{delta_f})
\cite{FurusakiHikihara,SirkerBortzJSTAT}. For $K<3/2$ ($\Delta>1/2$) the
integral is convergent and yields \cite{SirkerBortzJSTAT}
\begin{eqnarray}
F_B=\lambda\; {\rm Re}
[B(K+i\frac{Kh}{2\pi T},1-2K)]\l(\frac{2\pi T}{v}\r)^{2K-1}, \label{freeb}
\end{eqnarray}
where $B(x,y)=\Gamma(x)\Gamma(y)/\Gamma(x+y)$. For $K>3/2$ the integral
(\ref{delta_f_thermo}) is divergent and a cutoff has to be introduced
again. Then, the cutoff independent part of the integral is still given by
(\ref{freeb}) whereas the cutoff dependent terms are regular in
$h,T$ \cite{SirkerBortzJSTAT}. In the following discussion we will neglect
these regular terms. From (\ref{delta_f_thermo}) the boundary spin
susceptibility is easily obtained
\begin{eqnarray}
\fl\chi_{\rm B}&=&\biggl.-\frac{\partial^2  F_{B}}{\partial h^2}
\biggr|_{h=0} = - \lambda \l(\frac{K}{v}\r)^2
B(K,1-2K)[\pi^2-2\psi'(K)]\l(\frac{2\pi T}{v}\r)^{2K-3}, 
\label{chi1}
\end{eqnarray}
with $\psi'(x)=d\psi(x)/dx$. 
Note that for $1<K<3/2$ ($1/2<\Delta<1$), the boundary spin
susceptibility $\chi_{\rm B}$ shows a divergent behaviour $\sim 1/T^{3-2K}$,
as temperature decreases.
This anomalous temperature dependence is also observed in the boundary part 
of the specific heat coefficient given by,
\begin{eqnarray}
\fl\frac{C_{\rm B}}{T}&=&-\frac{\partial^2 F_{B}}{\partial T^2} 
=\lambda \l(\frac{2\pi}{v}\r)^2 (2K-1)(2K-2)B(K,1-2K)\l(\frac{2\pi T}{v}\r)^{2K-3}.
\label{heat1}
\end{eqnarray}
We would like to stress that in the formulas (\ref{chi1}) and (\ref{heat1})
there is no free parameter, and the pre-factors are exactly obtained.  These
divergent behaviours are physically understood as follows.  In contrast to the
bulk Heisenberg chain in which the ground state is a spin singlet state, spin
singlet formation in the vicinity of boundaries is strongly disturbed. For the
susceptibility, for example, we can write $\chi_B = \int_0^\infty \chi^u_B(x)$
where $\chi^u_B(x)$ follows from (\ref{delta_f_thermo}). Note, that the local
boundary susceptibility has also an alternating part \cite{EggertAffleck95},
which, however, does not contribute to $\chi_B$. A detailed comparison of
numerical data for the local boundary susceptibility (consisting of the
uniform and the alternating part) and field theory has been presented in
Ref.~\cite{BortzSirker}. In Fig.~\ref{Fig_chiB_local}, $\chi^u_B(x)$ is
plotted for various temperatures showing that the spins near the boundary are
more susceptible than the spins in the bulk. It should, however, be emphasised
that the singular behaviours are not due to the presence of boundary
operators, but interpreted as a consequence of finite-temperature corrections
%% of the surface energy and the boundary entropy $\ln\langle 0|B\rangle$ 
caused by bulk irrelevant interactions.

At zero temperature with a finite magnetic field,
a similar singular behaviour appears in the field dependence of
the boundary spin susceptibility given by,
\begin{eqnarray}
\chi_{\rm B}(T=0)= \lambda\l(\frac{K}{v}\r)^2 \sin(\pi K)
\Gamma(3-2K)\l(\frac{Kh}{v}\r)^{2K-3}. \label{zerospin}
\end{eqnarray}
The zero temperature susceptibility can also be derived from the Bethe ansatz
exact solution by using the Wiener-Hopf method. This has been done in
Ref.~\cite{BortzSirker,SirkerBortzJSTAT} and agrees with the expression
above.

\section{The isotropic case $\Delta = 1$}
\label{IsoCase}
At the isotropic point Umklapp scattering becomes marginal. Instead of
using (\ref{int1}) one can write the marginal perturbation in a
manifestly SU(2) invariant form using the SU(2) current operators. The
perturbation in this formulation becomes
\begin{equation}
\delta H =\int\frac{dx}{2\pi}[g_{\parallel}J_z\bar{J}_z
+\frac{g_{\perp}}{2}(J_{+}\bar{J}_{-}+J_{-}\bar{J}_{+})].
\label{int_iso}
\end{equation}
The small running coupling constants $g_\parallel$, $g_\perp$ fulfil
a set of known Renormalisation Group (RG) equations \cite{Lukyanov,Affleck98}
and are in general functions of the different scales, $T,L$ and $h$,
involved. The perturbation (\ref{int_iso}) can be split into two
parts: $J_z\bar{J}_z \propto \Pi^2-(\partial_x\phi)^2$ and
$J_{+}\bar{J}_{-}+J_{-}\bar{J}_{+} \propto \cos(\sqrt{8\pi}\phi)$. The
first term can be absorbed in the free electron Hamiltonian (\ref{H0})
by a rescaling of the fields $\Pi$ and $\phi$.  This leads to a
renormalisation of the Luttinger parameter $K\to
1+g_\parallel/2+\mathcal{O}(g^2)$ (see \ref{app_B} for details). With
this replacement in (\ref{scalechi}) we see that the susceptibility at
the isotropic point in the thermodynamic limit is now given by
\begin{equation}
\label{bulk_iso}
\chi_\bulk(T) = \frac{1}{\pi^2} \l( 1+ \frac{g(T)}{2}+\mathcal{O}(g^2)\r) \; ,
\end{equation}
a result first derived in Ref.~\cite{egg94}. The second term leads in
lowest order to a replacement $\lambda\to -g_\perp/4$. More generally,
the results for the anisotropic model derived in the previous sections
have to be expanded in powers of $1-1/K$ and then re-expressed in
terms of the coupling constants $g_\parallel$ and $g_\perp$
\cite{Lukyanov}. For details the reader is referred again to
\ref{app_B}. Here we only want to give the main results.

First, we consider the isotropic limit of our first order result for
the boundary susceptibility (\ref{chi1}). This leads to
\begin{equation}
\label{chiB_iso}
\chi_B^{(1)} = -\frac{g_\perp}{12T}
-\frac{g_\parallel g_\perp}{8T}\l(\frac{1}{2}-\frac{\Psi''(1)}{\pi^2}\r) \, .
\end{equation}
The second order result derived in \ref{app_A} also yields a
contribution in quadratic order in $g$ 
\begin{equation}
\label{chiB_iso2}
\chi_B^{(2)} \approx \frac{g_\perp^2}{8T}\cdot 0.11 \; .
\end{equation}
Here the factor $0.11$ stems from a numerical evaluation of the
integral in (\ref{A9}) for $K=1$. With $g_\parallel\to g$,
$g_\perp\to -g$ we therefore obtain up to quadratic order in $g$
\begin{equation}
\label{chiB_iso3}
\chi_B = \frac{a}{\pi^2}+\frac{g}{12T}+\frac{g^2}{8T} \l(0.66 -
\frac{\Psi''(1)}{\pi^2}\r) + \mathcal{O}(g^3) \, .
\end{equation}
Here $g=g(T)$ and is given by \cite{Lukyanov}
\begin{equation}
\label{coup_iso}
1/g + \ln(g)/2 = \ln\l(T_0/T\r) \; ,
%% \tilde{\lambda}_1(T,\delta E)
%% =\frac{1}{4}\ln^{-1}\l(\frac{\Lambda}{\mbox{max}(T,\delta E)}\r)
\end{equation}
with $T_0= \sqrt{\pi/2} e^{1/4+\gamma}$ where $\gamma$ is the Euler
constant. Note, that the scale in the logarithm, $T_0$, is chosen
according to the expansion of the bulk susceptibility. $T_0$ is
non-universal and can be fixed differently to achieve the best
possible agreement with the expansion of the boundary susceptibility
at the isotropic point. However, this is not important for $\chi_B$ as
it only influences sub-leading terms so we keep the scale as set by
the bulk part in order to have a single $g$ in the formula for
$\chi(L,T)$ discussed in the following. A more detailed discussion of
this point is presented in \ref{app_B}.

\begin{figure}[!htp]
\begin{center}
\includegraphics*[width=0.9\columnwidth]{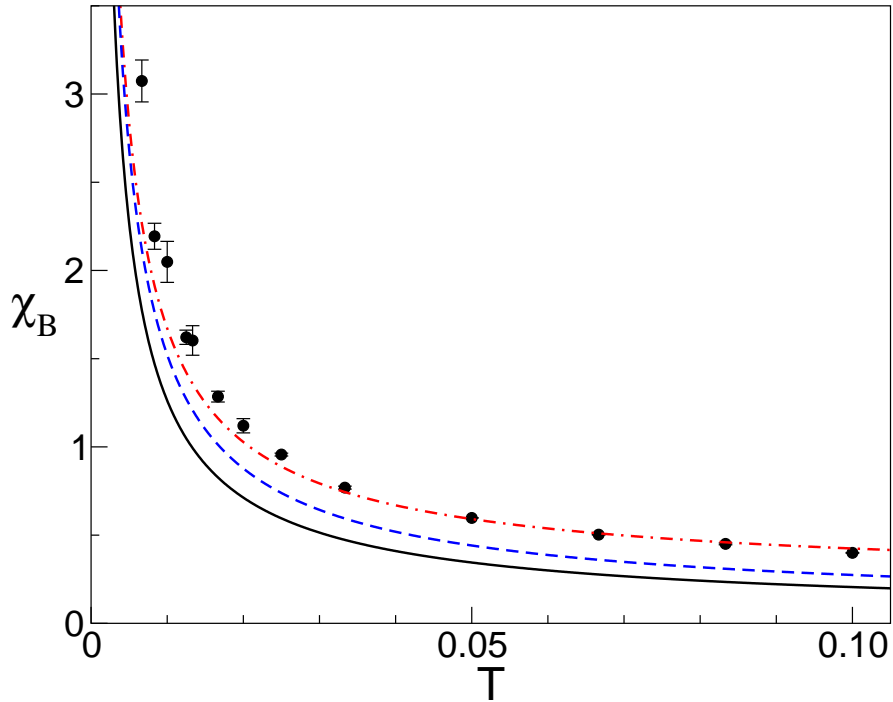}
\caption{Comparison between QMC results (dots) and the field theory result
  (\ref{chiB_iso3}). The black solid line corresponds to taking into account
  the first order in $g$ term only, the blue dashed line is the result with
  the second order term included, and the red dot-dashed line is obtained by
  adding a constant $Ka/2\pi v$ with $a=1.5$ ($K=1$, $v=\pi/2$). The QMC data
  for $\chi_B$ have been obtained by subtracting the bulk susceptibility known
  from Bethe ansatz \cite{egg94} from numerical data for $\chi(L)$. The chain
  length $L$ has always been chosen such that $TL/v \gtrsim 4$. In addition,
  the susceptibility has been calculated for lengths $L$ and $L+1$ and then
  averaged to further reduce finite size effects (see also
  Eq.~(\ref{deltaChiB}) and Fig.~\ref{Fig_delta_ChiB} below).}
\label{Fig_chiimp}
\end{center}
\end{figure}

In addition, we have added a contribution $\sim Ka/(2\pi v)$ to
(\ref{chiB_iso3}) with some constant $a$ stemming from the boundary
operator as in the anisotropic case. However, the constant
(\ref{NonU3}) obtained by Bethe Ansatz in the anisotropic case does
have an accumulation point of singularities at $K=1$ making it
impossible to extract the value at the isotropic point from this
formula. If one considers, on the other hand, the integral for the
boundary susceptibility derived directly by Bethe Ansatz for the
isotropic case (see Eq.~(36) in \cite{BortzSirker}), this task is
difficult as well because the logarithmic contributions coming from
the cut along the imaginary axis completely dominate. Indeed, at
low-temperatures the constant contribution is less important than any
of the logarithmic terms coming from the expansion of $\chi_B$
outlined above. Nevertheless, this constant becomes important again at
higher temperatures (but still $T/J\ll 1$) as in the anisotropic
case. We will therefore use it as a fitting parameter when comparing
to the numerical data in Fig.~\ref{Fig_chiimp}.
Note, that $a$ now effectively also partly incorporates higher order
logarithmic corrections to (\ref{chiB_iso3}) as well as the constant
coming from the boundary operator. Furthermore, we must in principle
replace $Ka/(2\pi v)\to (a/\pi^2)(1+g/2+g^2/4+\cdots)$ (see
\ref{app_B}). However, because the logs dominate at low temperatures
this merely leads to a small rescaling of $a$ which we will ignore
because $a$ is an ``effective constant'' in any case. Solving
Eq.~(\ref{coup_iso}) we can also write the boundary susceptibility as
\begin{equation}
\label{chiB_logs}
\chi_B =
\frac{1}{12T\ln(T_0/T)}\l(1-\frac{\ln\ln(T_0/T)}{2\ln(T_0/T)}+\cdots\r) \; .
\end{equation}
This means that even in the limit $T/v\gg 1/L$ we find a Curie-like
contribution albeit with a ``Curie constant'' which depends
logarithmically on temperature.

\begin{figure}[!htp]
\begin{center}
\includegraphics*[width=0.9\columnwidth]{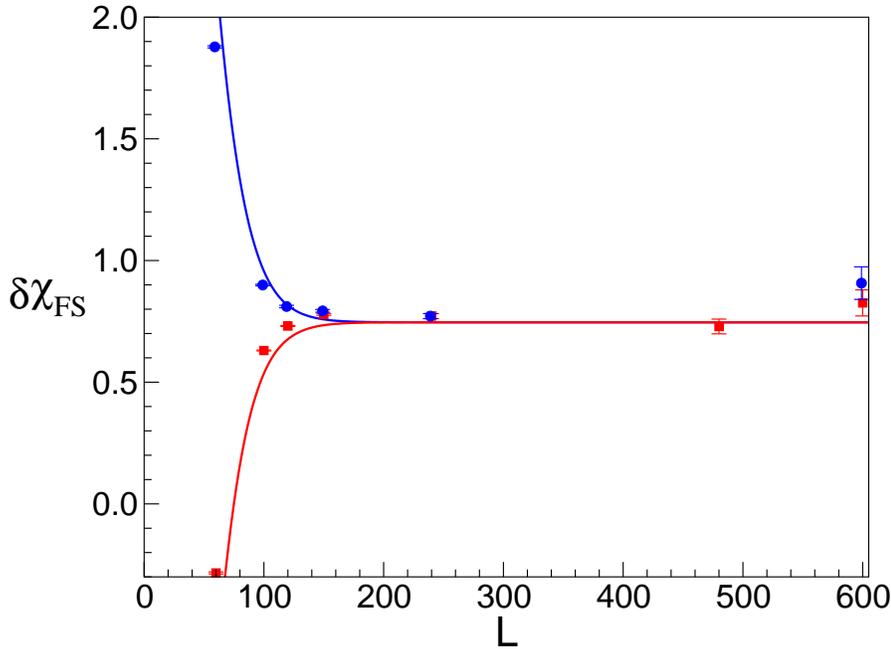}
\caption{QMC data for the susceptibility $\delta\chi_{FS}$ at temperature
  $T=J/30$ as a function of chain length. These data have been obtained by
  subtracting the bulk susceptibility known from Bethe ansatz \cite{egg94}
  from numerical data for $\chi(L)$. The red squares (blue circles) correspond
  to even (odd) chain length, respectively. The lines represent the
  approximation (\ref{deltaChiB}).}
\label{Fig_delta_ChiB}
\end{center}
\end{figure}

If $T/v$ becomes of order $1/L$ the susceptibility can no longer be
split into a bulk and a boundary contribution. However, we can still
subtract the known bulk contribution and define a quantity
$\delta\chi_{FS}$ as in Eq.~(\ref{deltaf}). In a first approximation,
we can use the thermodynamic limit result for $\chi_B$,
Eq.~(\ref{chiB_iso3}), and add the leading finite size corrections
stemming from the scaling part (second term in the last line of
Eq.~(\ref{limits})). This leads to
\begin{equation}
\label{deltaChiB}
\delta\chi_{FS}(L,T) \approx \chi_B(T) \pm \frac{8L^2T}{\pi^2}(1+g)\exp[-2LT(1+g/2)]
\end{equation}
where the plus (minus) sign refers to chains of odd (even) length and
we have replaced $K\to 1+g/2$ with $g$ determined by
Eq.~(\ref{coup_iso}). A comparison between this formula and QMC data
is shown in Fig.~\ref{Fig_delta_ChiB}.
Note, that Eq.~(\ref{deltaChiB}) implies an almost perfect cancelation of finite
size corrections when calculating the susceptibility for lengths $L$ and $L+1$
and then taking the average. This property has been used to improve the
numerical data for $\chi_B$ presented in Fig.~\ref{Fig_chiimp}.

% If $T/v\sim 1/L$ one might in a first approximation still use the
% thermodynamic limit result for $\chi_\bulk$ and $\chi_B$ and add the
% leading finite size corrections stemming from the scaling part (second
% term in the last line of Eq.~(\ref{limits})). If we define 

To obtain the result for $\chi(L,T)$ at the isotropic point in
general, we have to consider both parts of the perturbation
(\ref{int_iso}): First, we have to replace $K\to 1+g/2$ in the
exponentials of (\ref{scalechi}). For the correction (\ref{suscicorr})
we can only obtain a result to first order in $g$ because we do not
know the analytical solution of the integral and therefore cannot
easily expand it around $K=1$. The result to first order in $g$ is
readily obtained by just replacing $\lambda\to g/4$ and evaluating the
integral for $K=1$. Now, however, the running coupling constant $g$
depends on two scales: the length $L$ and temperature $T$. There is no
general solution of the renormalisation group equations if two
different scales are involved.  At low enough energies, however, the
smaller length scale will always dominate and the running coupling
constant becomes \cite{Lukyanov}
\begin{equation}
\label{coup_iso2}
1/g + \ln(g)/2 = \ln\l(\sqrt{2/\pi} e^{1/4+\gamma}\mbox{min}[L,v/T]\r) \; .
%% \tilde{\lambda}_1(T,\delta E)
%% =\frac{1}{4}\ln^{-1}\l(\frac{\Lambda}{\mbox{max}(T,\delta E)}\r)
\end{equation} 
In addition, the constant $a$ stemming from the boundary operator is
still present and we will use it again as a fitting parameter. Note,
that $a$ now will be different from the one obtained by fitting the
numerical data for $\chi_B$ as it also partly incorporates the
logarithmic terms now taken into account only to lowest order.

\begin{figure}[!htp]
\begin{center}
\includegraphics*[width=0.9\columnwidth]{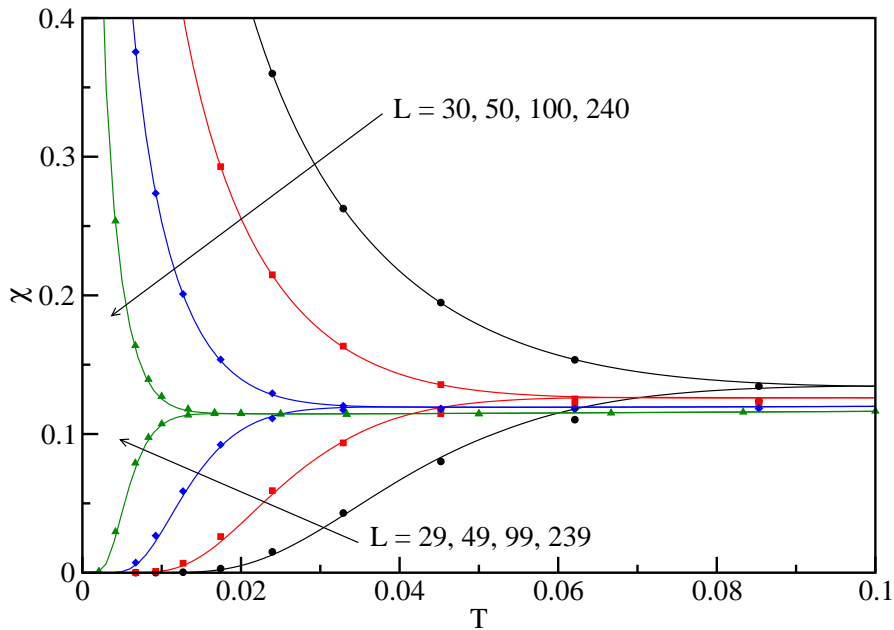}
\caption{Comparison for $\Delta=1$ between QMC results (symbols) and
  the field theory result (lines) with $a=4$ for chains of different
  length $L$.}
\label{Fig_iso_compare}
\end{center}
\end{figure}

To summarise, the susceptibility for the isotropic case is given by
$\chi(L,T) = \chi_0+\delta\chi_1$ with the replacements $K\to
1+g(L,T)/2+a/L$ in the exponentials of (\ref{scalechi}) and
$\lambda\to g(L,T)/4$ in (\ref{suscicorr}). A comparison between this
formula and QMC data for chains of different lengths is shown in
Fig.~\ref{Fig_iso_compare}.
Here we find that $a\approx 4$ works best describing the numerical
data over a wide temperature range and for different even and odd
chain lengths very accurately.\footnote{In \cite{SirkerLaflorencie} we
  used $a=5.8$. This difference is due to the fact that in this
  article we expanded in $a$, i.e., we used (\ref{NonU}) as well as a
  second order solution for (\ref{coup_iso}). Here we put $a$ directly
  into the exponentials of (\ref{scalechi}) and solve (\ref{coup_iso})
  numerically.}

The disadvantage of this formula for $\chi(L,T)$ is, that the integral
in (\ref{suscicorr}) has to be evaluated numerically for each length
and temperature considered. In the next section we will derive a much
simpler formula in the limit $T/J\ll 1/L$. We find empirically that
this formula describes the susceptibility over a much larger range of
temperatures than anticipated making it useful to fit experimental
data without the necessity to evaluate the integral in
(\ref{suscicorr}) and without having a free fit parameter.

\section{The ground state limit: $LT/v\to 0$}
\label{ground_state}
Next, we want to consider the limit $LT/v\to 0$. We have already shown
in Eq.~(\ref{limits}) that the scaling part of the susceptibility
shows a Curie-like divergence if the chain length is odd and an
exponentially suppressed behaviour if it is even. For the first order
correction (\ref{suscicorr}) we will concentrate again on the case
$K<3/2$ where the cutoff can be dropped. In the limit $LT/v\to 0$ we
find $\eta\l(\e^{-\frac{\pi v}{TL}}\r) \to \e^{-\frac{\pi v}{24TL}}$,
$\theta_1\l(\pi y,\e^{-\frac{\pi v}{2TL}}\r) \to 2 \e^{-\frac{\pi
    v}{8TL}}\sin(\pi y)$ and
\begin{equation}
 g_0\l(y,\e^{-\frac{\pi v}{KTL}}\r)\to \l\{\begin{array}{cc}
  0 & \mbox{L odd} \\[0.3cm]
  -2\cos(4\pi y)\e^{-\frac{\pi v}{KTL}} + 2\e^{-\frac{\pi v}{KTL}}   & \mbox{L
  even}
\end{array}
\r.
\end{equation} 
Here we have expanded $g_0$ only to first power in the small parameter
$\e^{-\frac{\pi v}{KTL}}$. For $L$ odd the first non-vanishing contribution is
second order. As the scaling part shows a power-law divergence in this case,
any exponentially small corrections can be safely neglected in any case. For
$L$ even, on the other hand, they are important as the scaling part is also
exponentially small. In the even case we can now evaluate the integral and
find
 \begin{equation}
 \delta\chi_1(LT/v\to 0) = \l\{\begin{array}{cc}
  0 & \mbox{L odd} \\[0.3cm]
  \frac{2\lambda}{T^2}\l(\frac{\pi}{L}\r)^{2K}\frac{\Gamma(3-2K)}{\Gamma(2-K)\Gamma(3-K)}\e^{-\frac{\pi
  v}{KTL}}& \mbox{L
  even}
\end{array}
\r.
\end{equation}
Finally, lets consider the contribution $\delta\chi_2$. For $L$ odd
$g_2(\e^{-\frac{\pi v}{KLT}})\to 0$ in lowest order whereas
$g_2(\e^{-\frac{\pi v}{KLT}})\to 2\e^{-\frac{\pi v}{KLT}}$ for $L$ even. This
leads to 
\begin{equation}
 \delta\chi_2 = \l\{\begin{array}{cc}
  0 & \mbox{L odd} \\[0.3cm]
  \frac{2\pi v}{K T^2L^3}\e^{-\frac{\pi v}{KTL}}\cdot a   & \mbox{L
  even}
\end{array}
\r.
\end{equation}
For the susceptibility in the limit $LT/v\to 0$ we therefore find
\begin{equation}
\label{our_formula}
 \chi(LT\to 0) = \l\{\begin{array}{cc}
  \frac{1}{4LT} & \mbox{L odd} \\[0.3cm]
  \frac{2}{LT}\e^{-\frac{\pi
  v}{KTL}}\l[1+\lambda\frac{\beta \pi^{2K}}{L^{2K-1}T}+\frac{\pi
  v}{KL^2T}a\r] & \mbox{L
  even}
\end{array}
\r.
\end{equation}
where $\beta = \Gamma(3-2K)/[\Gamma(2-K)\Gamma(3-K)]$. So corrections to the
scaling limit result in first order in $\e^{-\frac{\pi v}{TL}}$ are only
present in the case of even chain length. However, even in this case these
corrections are suppressed by additional powers of $1/(LT)$. So one might
think that Eq.~(\ref{our_formula}) is of purely academic interest. However, as
we will show below, it enables us to derive a simple formula for the
susceptibility of the isotropic chain which works over a much larger range of
temperatures and lengths than anticipated.

In the isotropic case we have to replace $K\to 1+g_\parallel(L)/2$ in
the scaling part (\ref{scalechi}).
%% where $g(L)$ is given by (\ref{coup_iso}) with $L$ being the
%% relevant scale in this limit. 
Furthermore $\lambda\to -g_\perp(L)/4$, so that (\ref{our_formula}) at
the isotropic point reads
\begin{equation}
\label{our_formula_iso}
\fl \chi(LT\to 0) = \l\{\begin{array}{cc}
  \frac{1}{4LT} & \mbox{L odd} \\[0.3cm]
 \frac{2}{LT}\e^{-\frac{\pi^2}{2TL}}\l[1+\frac{\pi^2}{4LT}g_\parallel(L)-\frac{\pi^2}{4LT}g_\perp(L)+\frac{\pi^2}{2L^2T}a\r] & \mbox{L
  even}
\end{array}
\r.
\end{equation} 
where $v=\pi/2$ has been employed. Note that with $g_\parallel \to g$,
$g_\perp \to -g$ the replacement $K\to 1+g_\parallel(L)/2$ in the
scaling part and the first order contribution in $\lambda\to
-g_\perp(L)/4$ both produce exactly the same correction
$\frac{\pi^2}{4LT}g(L)$ in the brackets in
(\ref{our_formula_iso}). Here $g(L)$ is given by (\ref{coup_iso}) with
$L$ being the relevant scale in this limit. We want to stress that
both terms yielding exactly the same contribution in the limit
considered here contribute very differently in the thermodynamic
limit. Whereas the $g_\parallel$-part leads to a correction to the
bulk susceptibility, the $g_\perp$-part determines the boundary
susceptibility. 

Ignoring the contribution $\sim a$ which comes from the boundary
operator and is suppressed by an additional power of $L$ we can
therefore obtain exactly the same expansion by replacing $1/K\to
1-g(L)$ in the exponentials of (\ref{scalechi}). I.e., we can absorb
the first order correction $\delta\chi_1$ into the scaling form. This
leads to the formula
\begin{equation}
\label{Seb-formula}
\chi =\frac{1}{LT} \frac{\sum_{S_z} S_z^2 \exp\l[-\frac{\pi^2}{2LT}(1-g(L))S_z^2\r]}{\sum_{S_z} \exp\l[-\frac{\pi^2}{2LT}(1-g(L))S_z^2\r]} \, .
\end{equation}
This formula should be valid in the limit $LT/v\to 0$ when only the ground
state and the lowest excited states contribute to the sum over $S_z$. One
easily verifies that Eq.~(\ref{our_formula_iso}) is indeed reproduced in this
limit: For $L$ odd we take only the states $S_z=\pm 1/2$ into account and the
correction $1-g(L)$ cancels out. For $L$ even we consider the ground state
$S_z=0$ and the first excited state $S_z=\pm 1$. Expanding in $g(L)$ then
leads to (\ref{our_formula_iso}).

So starting from our general result for the susceptibility of finite open
chains we have found that the correction to the excitation energy of the
lowest excited states with $S_z\pm 1$ in the isotropic case is given by $E\to
(\pi v/L)[1-g(L)]$. This agrees with the findings in Ref.~\cite{AffleckQin}.
Corrections to higher excited states are not of such simple form, and, in
general, it is therefore not possible to absorb these corrections into the
scaling part.

\begin{figure}[!htp]
\begin{center}
\includegraphics*[width=0.9\columnwidth]{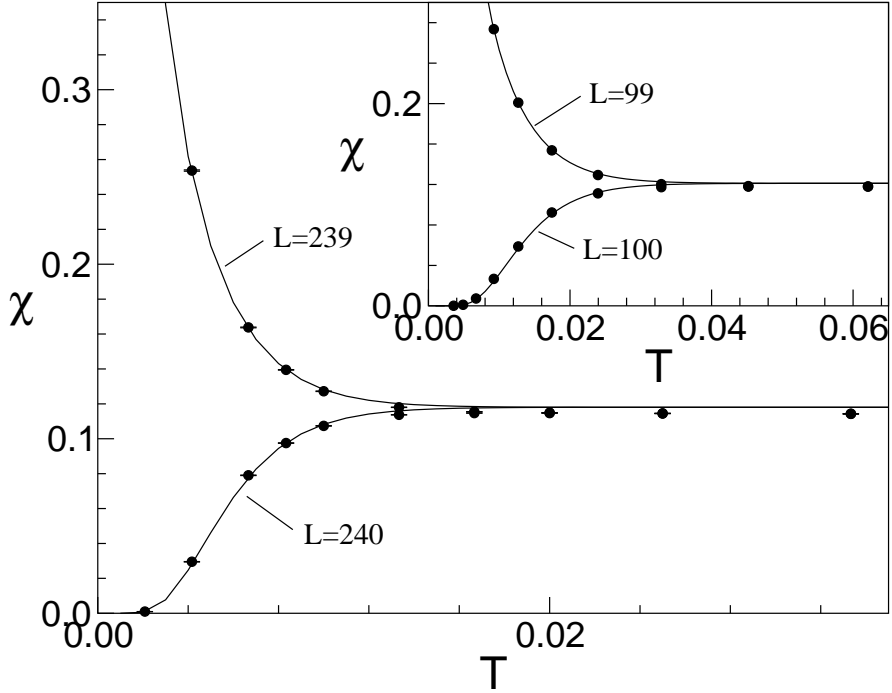}
\caption{Comparison for $\Delta = 1$ between QMC results (dots) and
  the field theory result (\ref{Seb-formula}) in the limit $LT/v\to 0$
  (lines).}
\label{Fig_comp_SFormula}
\end{center}
\end{figure}

Empirically, however, we find that (\ref{Seb-formula}) works quite well over a
much larger range of lengths and temperatures than one might anticipate. In
Fig.~\ref{Fig_comp_SFormula} we compare this formula again to QMC data.
Although the agreement at higher temperatures where the susceptibility
becomes constant is not perfect, the errors in this region are only
about $3\%$.
%% For most practical purposes, in particular
%% to calculate the average susceptibility $\chi_p$ for impurity concentration
%% $p$ we can therefore use (\ref{Seb-formula}) which is relatively easy to
%% evaluate. 

\section{The averaged susceptibility and the effective Curie constant}
\label{avg_susci}
In this section we want to discuss the averaged susceptibility of the
isotropic Heisenberg chain for a concentration of chain breaks $p$ as
defined in Eq.~(\ref{chi_p}). Here we have assumed that we have a
completely random (Poisson) distribution of impurities. However, we
could obtain $\chi_p$ for any other distribution as well. $\chi_p$ can
be calculated by using our field-theory results in section
\ref{IsoCase} for $L\gtrsim 10$ and $T/J\lesssim 0.1$ in combination
with exact diagonalisation or QMC results for shorter chains. This
method has been used in Ref.~\cite{SirkerLaflorencie} to analyse
susceptibility data for Sr$_2$Cu$_{1-x}$Pd$_x$O$_{3+\delta}$
\cite{Kojima}. Based on our analysis of the limits $LT/v\to 0$ and
$LT/v\to\infty$ in the previous sections we want to derive here a
simple formula which allows to determine the concentration of chain
breaks by fitting the measured susceptibility.

We have seen that there are no corrections to the scaling limit result in the
limit $LT/v\to 0$ for odd chains. For even chains the susceptibility is
exponentially small in this limit in any case so that even chain segments
practically do not contribute to the average in (\ref{chi_p}). In the opposite
limit we can write $\chi(L)=L\chi_\bulk +\chi_B$. We now assume
that the crossover occurs at some length $L_c=\gamma J/T$ where $\gamma$ is a
crossover parameter which we expect to be of order one. For the average
susceptibility we can therefore write
\begin{eqnarray}
\label{fit_formula}
\fl \chi_p &\approx& \frac{p^2}{4T}\sum_{L\; \mbox{\tiny odd}}^{L_c}(1-p)^L +
p^2\sum_{L_c}^\infty (L\chi_\bulk+\chi_B) (1-p)^L \\
\fl &=& \frac{p}{4T}\frac{1-p}{2-p}\l(1-(1-p)^{\gamma/T}\r)+(1-p)^{\gamma/T}\l[\l(1-p+\frac{p\gamma}{T}\r)\chi_\bulk+p\chi_B\r] \nn
\end{eqnarray}
and we use the field theory result for the bulk susceptibility \cite{Lukyanov}
\begin{equation}
\label{bulk_thermo}
\chi_\bulk = \frac{1}{\pi^2}\l(1+\frac{g(T)}{2}+\frac{3g^3(T)}{32}+\sqrt{3}{\pi}T^2\r) 
\end{equation}
where the $T^2$-term stems from the irrelevant operators with scaling
dimension $4$. The boundary susceptibility $\chi_B$ is given by
(\ref{chiB_iso3}) with $a=1.5$ as in Fig.~\ref{Fig_chiimp}.
\begin{figure}[!htp]
\begin{center}
\includegraphics*[width=0.9\columnwidth]{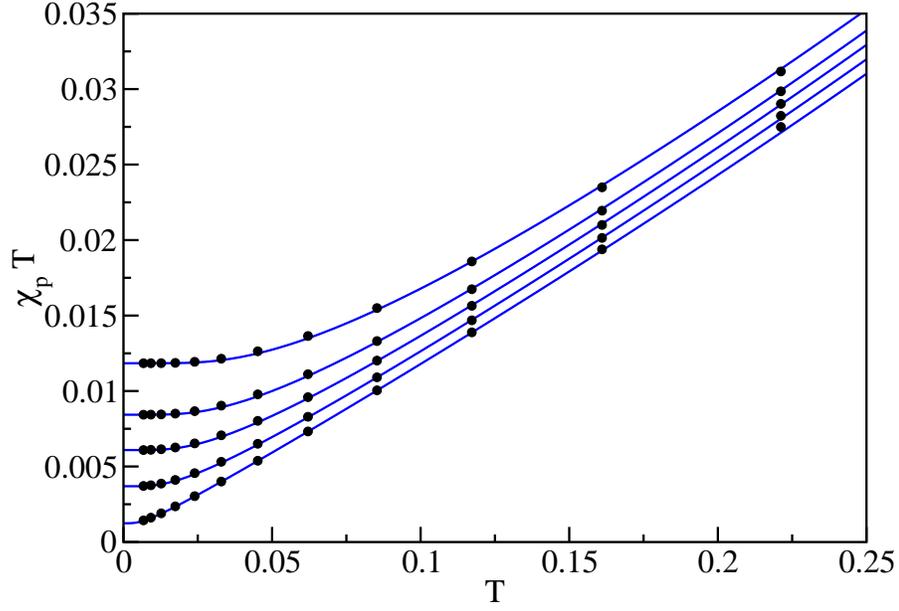}
\caption{$T\chi_p$ for $p=0.01,0.03,0.05,0.07,0.1$ from bottom
  to top. The black dots are obtained by using QMC data for $\chi(L)$ and the
  lines represent the fitting formula (\ref{fit_formula}) with
  $\gamma=1$.}
\label{Fig_AvgSusci}
\end{center}
\end{figure}
A comparison between QMC data for $\chi_p$ and (\ref{fit_formula})
with $\gamma=1$ for various impurity concentrations is shown in
Fig.~\ref{Fig_AvgSusci}.

\begin{figure}[!htp]
\begin{center}
\includegraphics*[width=0.9\columnwidth]{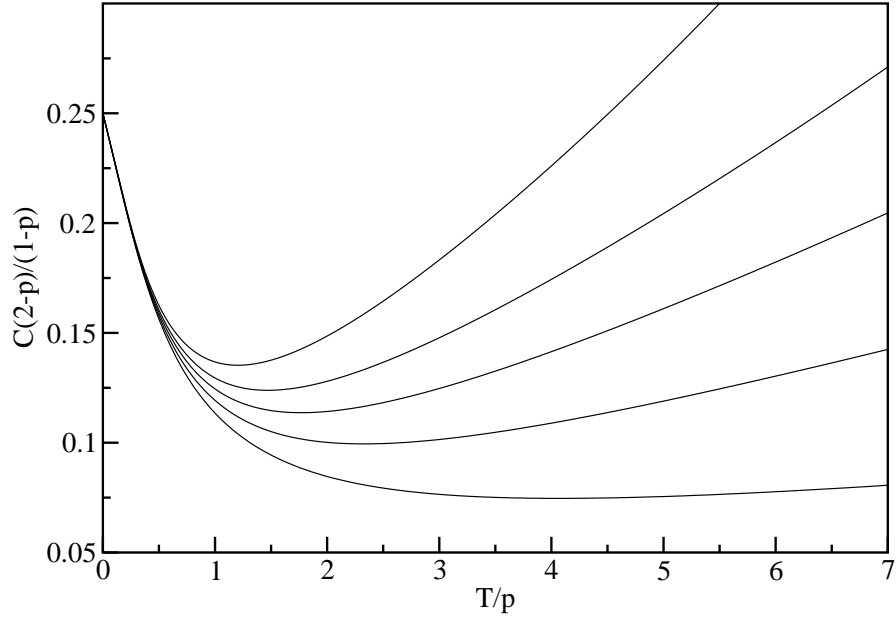}
\caption{The effective Curie constant as defined in Eq.~(\ref{Curie_constant})
  for $p=0.01,0.03,0.05,0.07,0.1$ (from bottom
  to top) as a function of $T/p$.}
\label{Fig_Curie}
\end{center}
\end{figure}

The impurity part and an effective Curie constant $C$ can then be
defined according to Eq.~(\ref{chi_p}) by
\begin{equation}
\label{Curie_constant}
p\delta\bar{\chi}_{FS} =\chi_p-(1-p)\chi_\bulk =p\frac{C}{T} \; .
\end{equation}
This effective Curie constant is shown in Fig.~\ref{Fig_Curie}.
Note, that only in the extreme low-temperature limit the usual Curie
constant $C\to 1/4\cdot (1-p)/(2-p)\approx 1/8$ (half of the chains
contribute a Curie constant of $1/4$ for $p\ll 1$) is
recovered. However, $C(2-p)/(1-p)$ as a function of $T/p$ shows an
almost perfect collapse (scaling) for different impurity
concentrations if $T/p\lesssim 1/2$. At higher temperatures this
scaling no longer holds because the boundary contribution $\chi_B$ is
not a scaling function in $T/p$. The non-trivial temperature
dependence of the effective Curie constant, in particular the finite
temperature minimum, and the scaling at temperatures $T\lesssim p/2$
provide a way to test this scenario experimentally.

\section{Inter- and Intrachain couplings}
\label{seg-couplings}
\begin{figure}[!htp]
\begin{center}
\includegraphics*[width=0.9\columnwidth]{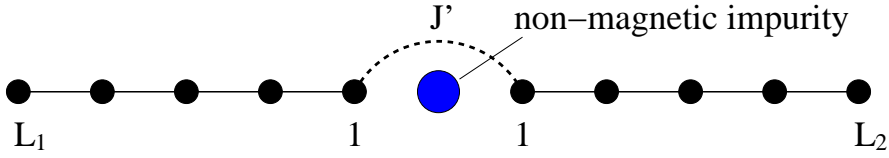}
\caption{Two chain segments with length $L_1$ and $L_2$, respectively, coupled
by a weak intrachain coupling $J'\ll J$.}
\label{Fig_2chains}
\end{center}
\end{figure}

In any real system there are residual couplings between the chain
segments in the principal chain direction (intrachain couplings). In
addition, there are also perpendicular couplings between the chains
(interchain couplings) which at low enough temperatures will destroy
the one dimensionality of the system and usually induce some sort of
magnetic order. In Sr$_2$CuO$_{3}$, for example, the nearest-neighbour
Heisenberg coupling along the chain direction (b-axis) is estimated to
be $J\sim 2200$ K whereas the couplings along the other axis are
orders of magnitude smaller ($J_a\sim 5$ K, $J_c\sim 10^{-3}$ K)
\cite{MotoyamaEisaki}. Furthermore, it has been pointed out that the
next-nearest-neighbour coupling along the chain direction is probably
not that small, $J_2\sim 140$ K \cite{RosnerEschrig}. This irrelevant
coupling will only cause small corrections to the susceptibility of an
isolated chain segment by changing the marginally irrelevant coupling
constant $\lambda$. However, it induces a coupling between different
chain segments even when the nearest-neighbour exchange $J$ is absent
due to a non-magnetic impurity. Note, that there is no frustration in
this case and $J_2$ just acts as an effective coupling $J'$ between
the chain ends as shown in Fig.~\ref{Fig_2chains}.
Although the next-nearest neighbour coupling across an impurity can
differ from the one across a magnetic ion, we will assume that it is
still of the same order of magnitude, i.e., $J'\sim J_2$.

The susceptibility of the weakly coupled chains is given by
\begin{equation}
\label{intra_coup1}
\chi = \frac{1}{LT}\l\langle \l(\sum_{i=1}^{L_1} S_{1i}^z + \sum_{j=1}^{L_2} 
S_{2j}^z\r)^2 \r\rangle 
\end{equation}
with $L=L_1+L_2$. Here $ S_{1i}^z$ ($S_{2j}^z$) denotes the
$z$-component of the spin on chain 1 (2) at site $i$ ($j$),
respectively. For $J'\ll J$ we can treat this coupling using
perturbation theory. In zeroth order this lead to $\chi^{(0)} =
(L_1\chi_1^{(0)} + L_2\chi_2^{(0)})/L$. In first order only the
longitudinal part of the coupling contributes and we obtain
\begin{eqnarray}
\label{intra_coup2}
\delta\chi &=& -\frac{2J'}{LT^2}\l\langle \sum_{i=1}^{L_1} S^z_{1i}
S^z_{11}\r\rangle\l\langle \sum_{j=1}^{L_2} S^z_{2j} S^z_{21}\r\rangle
\nonumber \\
&=& -\frac{2J'}{LT^2} \l\langle S^z_{1\mbox{\tiny tot}}
S^z_{11}\r\rangle\l\langle S^z_{2\mbox{\tiny tot}} S^z_{21}\r\rangle =  -\frac{2J'}{L}\chi_{11}\chi_{21}
\end{eqnarray}
where $\chi_{l1} =\partial < S^z_{l1} >/(\partial h)$ is the
susceptibility of the boundary spin $S^z_{l1}$ and $S^z_{l\mbox{\tiny
    tot}}=\sum_{i=1}^{L_l}S^z_{li}$ ($l=1,2$), respectively. The extra
factor of 1/T arises due to time-translational invariance and the
imaginary time integral in the first order perturbation theory
formula. Because of the broken spatial-translational invariance, the
local susceptibility is position dependent and the boundary spins are
more susceptible than the spins deep inside the chain.  These effects
have been studied in detail in Refs.~\cite{EggertAffleck95,
  BortzSirker}. As we are here just interested in an order of
magnitude estimate of the effect of intrachain coupling on $\chi_p$ we
can replace the susceptibility of the boundary spin $\chi_{l1}$ by the
average susceptibility per site $\chi_l^{(0)}$. This leads to
$\delta\chi \approx -2(J'/L)\chi_1^{(0)}\chi_2^{(0)}$. Assuming
further that for an impurity concentration $p$ all chain segments have
length $L\sim 1/p$ we find
\begin{equation}
\label{intra_coup3}
\delta\chi_p \sim J'p\chi_p^2 \; .
\end{equation}
At $T\ll pv$ we have $\chi_p\sim p/(8T)$. The correction
(\ref{intra_coup3}) therefore becomes of order $\chi_p$ at $T\sim J'
p^2$.\footnote{Note that there is a factor $p$ missing in
  Ref.~\cite{SirkerLaflorencie}.} For temperatures $T\gg J'p^2$ intrachain
coupling can therefore be neglected. 

Although first order perturbation theory can give us the temperature scale
where intrachain coupling becomes important, it is not sufficient once this
scale is approached. This can be seen by a scaling analysis. We have
$S^z\sim\partial_x\phi$ with scaling dimension one. In $n$-th order
perturbation theory in the intrachain coupling $J'$, the longitudinal part of
the coupling therefore yields $\chi^{(n)}\sim (TL)^{-1}(J'/TL^2)^n$. We can
therefore write the susceptibility in scaling form
\begin{eqnarray}
\label{intra_coup4}
\chi &\sim&
\frac{1}{LT}\l(1+\frac{J'}{L^2T}+\frac{J'^2}{L^4T^2}+\frac{J'^3}{L^6T^3}+\cdots\r)
\; .
\end{eqnarray}
Replacing again $L\sim 1/p$ we find for the averaged susceptibility
\begin{eqnarray}
\label{intra_coup5}
\chi_p&\sim&
\frac{p}{T}\l(1+\frac{J'p^2}{T}+\l(\frac{J'p^2}{T}\r)^2+\l(\frac{J'p^2}{T}\r)^3+\cdots\r)
\; .
\end{eqnarray}
This means that at temperatures where intrachain coupling $J'$ can no longer
be neglected, $T\sim J' p^2$, perturbation theory in $J'$ breaks down. At
temperatures $T\ll J' p^2 < Jp$, on the other hand, we can obtain an effective
model by replacing all segments of odd length by $S=1/2$ spins. The couplings
between these effective $S=1/2$ will then be random with a very broad
distribution. According to Fisher \cite{Fisher_IRFP} we might therefore expect
that a system of weakly coupled chain segments renormalises at very low
temperatures to the infinite random fixed point. In any realistic system,
however, there will be also interchain couplings between the chain segments
preventing the system from reaching this fixed point.

\begin{figure}[!htp]
\begin{center}
\includegraphics*[width=0.9\columnwidth]{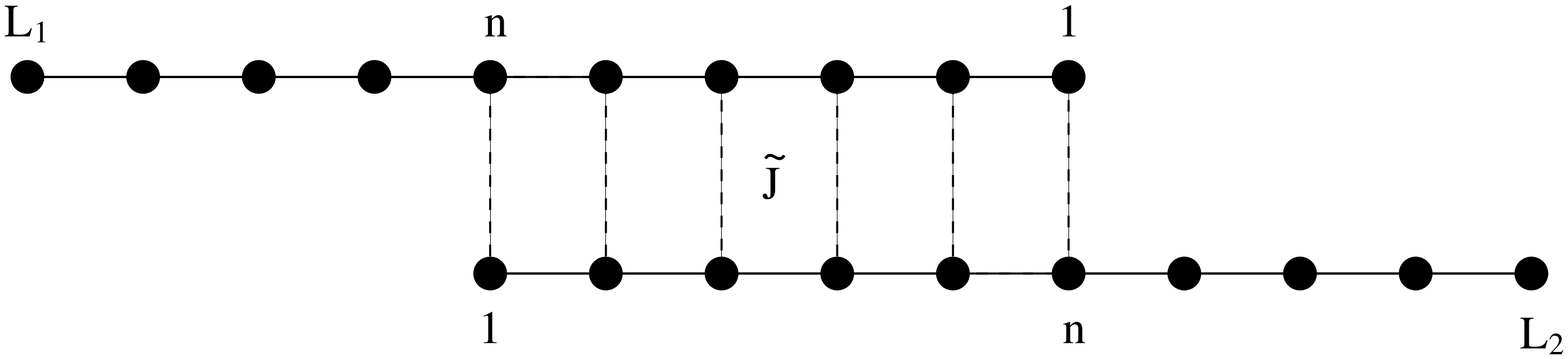}
\caption{Two chain segments with length $L_1$ and $L_2$, respectively, have an
  overlap of $n$ sites and are coupled by an interchain coupling $\tilde{J}$.}
\label{Fig_interchain}
\end{center}
\end{figure}

An interchain coupling between two chain segments is shown in
Fig.~\ref{Fig_interchain}.
In first order perturbation theory we obtain in this case
\begin{eqnarray}
\label{intra_coup6}
\fl\delta\chi^{(1)} &=& -\frac{2\tilde{J}}{LT^2}\sum_{k=1}^n\l\langle \sum_{i=1}^{L_1} S^z_{1,i}
S^z_{1,k}\r\rangle\l\langle \sum_{j=1}^{L_2} S^z_{2,j} S^z_{2,n+1-k}\r\rangle
\\
\fl &=& -\frac{2\tilde{J}}{LT^2} \sum_{k=1}^n\l\langle S^z_{1\mbox{\tiny tot}}
S^z_{1,k}\r\rangle\l\langle S^z_{2\mbox{\tiny tot}} S^z_{2,n+1-k}\r\rangle =
-\frac{2\tilde{J}}{L}\sum_{k=1}^n\chi_{1,k}\chi_{2,n+1-k} \nonumber
\end{eqnarray}
where $\chi_{l,k}$ denotes the local susceptibility of the $l$-th chain at site
$k$. To simplify matters, we consider the case of two chains with equal
lengths $L_1=L_2=L$ which completely overlap, i.e., $n=L$. In this case
Eq.~(\ref{intra_coup6}) becomes
\begin{equation}
\label{intra_coup6b}
\delta\chi^{(1)} = -\frac{2\tilde{J}}{L}\sum_{k=1}^L\chi_{1,k}\chi_{2,k} \: .
\end{equation}
At $T\ll pv$ we can again set $\chi_{1,k}\sim \chi_p\sim p/T$ which leads to
$\delta\chi^{(1)}_p \sim \tilde{J}p^2/T^2$. This becomes of order $\chi_p$ at
temperatures $T\sim \tilde{J}p$. As $\tilde{J}$ is of order of the N\'eel
temperature $T_N$ this scale is completely irrelevant. However, for a system
with open boundaries the local susceptibility has also a staggered part
\cite{EggertAffleck95,EggertAffleckHorton} which will contribute to
(\ref{intra_coup6b}). This is a consequence of the $S^z$-operator having a
staggered part, $S^z_{k,st.} \sim c(-1)^k \cos\sqrt{2\pi}\phi$, where $c$ is a
constant. For a chain with odd length in the ground state limit, $T\ll 1/L$,
we find \cite{EggertAffleckHorton}
\begin{eqnarray}
\label{intra_coup6c}
\fl \chi_k^{st.} &=& \frac{1}{T}\langle S^z_{tot} S^z_{k,st.}\rangle \approx
\frac{1}{T} \sum_{\pm} \langle \pm | S^z_{tot} S^z_{k,st.} | \pm \rangle \approx (-1)^k\frac{c}{T} \sqrt{\frac{\pi}{2L}\sin\frac{\pi k}{L}} 
\end{eqnarray} 
where $|\pm\rangle$ is the ground state with $S^z_{tot}=\pm 1/2$,
respectively. This leads to $\delta\chi^{(1)}\sim \tilde{J}/(LT^2)\sim
\tilde{J}p/T^2$ and therefore dominates compared to the contribution
originating from the uniform part of the local susceptibility. It becomes of
order $\chi_p$ at temperatures $T\sim \tilde{J}$.

Finally, we want to consider the ``clean limit'', $T\gg pv$. In this case, the
N\'eel temperature $T_N$ can be determined by treating the interchain coupling
in mean field theory. This leads to the well known condition
\begin{equation}
\label{intra_coup7}
z\tilde{J}\chi_{st}(T_N) = 1
\end{equation}
where $\chi_{st}$ is the staggered susceptibility, i.e., the staggered
response to the effective staggered field originating from the other
chains and $z$ is the coordination number.  Because $\chi_{st}(T)\sim
1/T$ we obtain that $T_N\sim \tilde{J}$. The same condition for the
breakdown of one-dimensionality in this limit is obtained by
considering the contribution of the staggered part of the
$S^z$-operator to the susceptibility in second order perturbation
theory in the interchain coupling.

We therefore conclude that our results for the averaged susceptibility
$\chi_p$ in the previous sections are applicable as long as $T\gg
\mbox{Min}(\tilde{J},p^2J')$.

\section{Experimental situation}
\label{experiment}
The best known realisation of the spin-$1/2$ Heisenberg chain is
Sr$_2$CuO$_3$. Here copper is in a 3d$^9$ configuration and has spin $1/2$.
The Heisenberg couplings between these spins are spatially very anisotropic as
already mentioned in the previous section. For $T\gg J_a\sim 5$ K interchain
coupling can therefore be neglected and the magnetic properties are described
by the one-dimensional isotropic Heisenberg model.
% They are coupled strongest along the crystallographic
% b-axis and the coupling constant has been estimated to be $J\equiv
% J_b\approx 2000-2400$ K \cite{who}. Couplings along the other axis are
% much weaker with $J_a\approx 5$ K and $J_c\approx 10^{-3}$ K
% \cite{who}. The next-nearest neighbour coupling along the b-axis has
% been estimated to be $J_2\approx 140$ K \cite{Roesner}. 

The doped compound Sr$_2$Cu$_{1-x}$Pd$_x$O$_3$ has been studied by Kojima {\it
  et al} \cite{Kojima}. Here palladium has spin zero and therefore serves as a
non-magnetic impurity. Impurity concentrations from $x=0.5\%$ up to $x=3\%$
have been studied. This means that $x^2J_2$ with $J_2\sim 140$ K is small even
compared to $J_a$. In this material the limit where our theory is applicable
is therefore set by the interchain coupling, $T\gg J_a\approx 5$ K.

The theoretical analysis of the susceptibility measurements on
Sr$_2$Cu$_{1-x}$Pd$_x$O$_3$ is hampered by the fact that even the pure
system Sr$_2$CuO$_3$ already shows Curie-like contributions at low
temperatures \cite{AmiCrawford,MotoyamaEisaki}. It has been observed
that these contributions can be significantly reduced by annealing and
two explanations have been offered: Excess oxygen might be present in
the as grown compound.  Most likely, each excess oxygen would then
dope two holes into the chain leading to two Zhang-Rice singlet type
states as in the high-$T_c$ compounds.  Assuming that these holes are
relatively immobile, they effectively act as chain breaks. In this
scenario the Curie-type contribution in the ``pure compound'' is
caused by the mechanism described in this paper. Indeed, it has been
shown \cite{SirkerLaflorencie} that the susceptibility data for the as
grown crystal sample in \cite{MotoyamaEisaki} can be well described by
assuming a concentration of chain breaks, $p=0.6\%$.  For the powder
samples studied by Ami {\it et al.}  \cite{AmiCrawford} an alternative
explanation was proposed by Hill {\it et al.} based on the observation
that Sr$_2$CuO$_3$ decomposes into Sr$_2$Cu(OH)$_6$ under exposure to
air and water \cite{HillJohnston}. The copper spins-$1/2$ in
Sr$_2$Cu(OH)$_6$ are only weakly coupled according to Hill {\it et
  al.} and can therefore induce a true Curie contribution as opposed
to the more complicated behaviour expected in the chain break
scenario. It is, however, hard to see how this could also explain the
Curie-like contributions observed on single crystals by Motoyama {\it
  et al.}  \cite{MotoyamaEisaki}. We therefore believe that the excess
oxygen scenario is the most plausible explanation in this case.

The Sr$_2$Cu$_{1-x}$Pd$_x$O$_3$ samples investigated by Kojima {\it et
  al.}~have not been annealed so that we expect chain breaks due to excess
oxygen in addition to the once caused by palladium. Furthermore, it can
certainly not be excluded that there are indeed additional true paramagnetic
impurity contributions caused by decomposition or other mechanisms. This makes
a thorough theoretical analysis very difficult. In \cite{SirkerLaflorencie}
such an analysis has been performed under the assumptions that true
paramagnetic Curie contributions can be neglected and that excess oxygen will
lead to immobile holes acting as additional chain breaks. The measured
susceptibility $\chi^{\mbox{\tiny exp.}}$ then consists of a temperature
independent part $\chi'$ due to core diamagnetism and Van Vleck paramagnetism
and $\chi_p$ coming from the Heisenberg chains with impurity concentration
$p$. Here $p=x+\delta$ has to be considered as a fitting parameter consisting
of the nominal palladium concentration $x$ and the concentration of additional
chain breaks $\delta$ due to excess oxygen. A best fit of the experimental
data was obtained in \cite{SirkerLaflorencie} with the two parameters $p$,
$\chi'$ as displayed in Table \ref{tab1}.
\begin{table}[htbp]
%% \squeezetable
  \caption{Concentration $x$ of Pd ions in experiment compared to impurity
    concentration $p$ and constant contribution $\chi'$ yielding the best
    theoretical fit. The first line corresponds to the ``as grown'' sample of Sr$_2$CuO$_{3}$ from
    Ref.~\cite{MotoyamaEisaki}.}
%%  as well as the minimum triplet gap for the columnar dimer state in
%%  Fig.~\ref{fig4}(a)}
\label{tab1}
\hspace*{0.2cm}
\begin{center}
\begin{tabular}{ccc} %{rdddd}
\multicolumn{1}{c}{$x$ (Exp.)} &\multicolumn{1}{c}{$p$ (Theory)} 
&\multicolumn{1}{c}{$\chi'$ [emu/mol]} \\
\hline\\[-0.3cm]
$0.0$ & $0.006$ & $-7.42\times 10^{-5}$ \\
$0.005$ & $0.012$ & $-7.7\times 10^{-5}$ \\
$0.01$ & $0.014$ & $-7.5\times 10^{-5}$ \\
$0.03$ & $0.024$ & $-7.5\times 10^{-5}$
\end{tabular}   
\end{center}                                                                                                   
\end{table}
For the sample with a nominal Pd concentration of $3\%$ our best
fields has yielded $p=2.4\%$, i.e., less than the nominal
concentration. One might speculate that the Pd ions cluster or that
part go in at interstitial positions instead of replacing Cu ions thus
reducing the number of chain breaks created.
 
\begin{figure}[!htp]
\begin{center}
\includegraphics*[width=0.9\columnwidth]{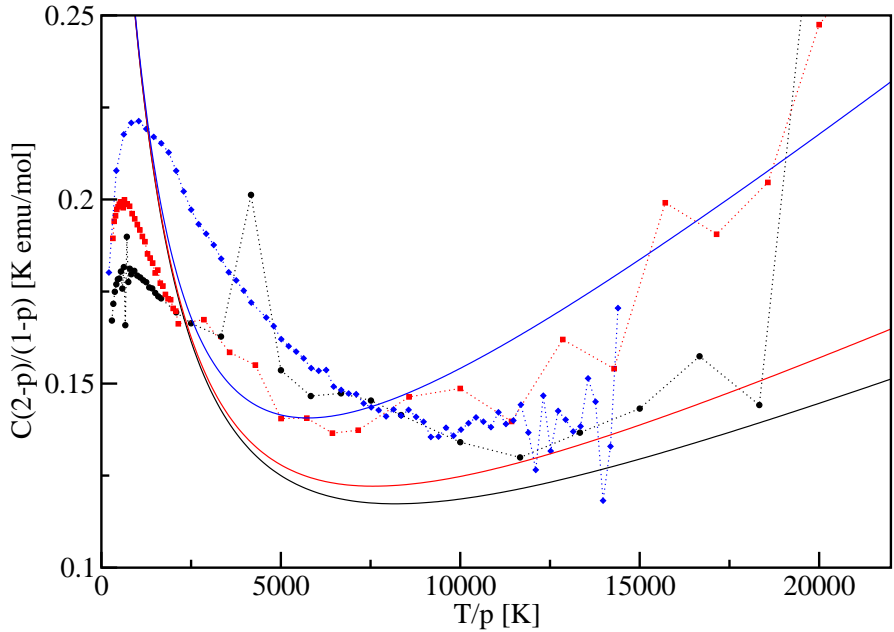}
\caption{The effective Curie constant (\ref{Curie_exp}) extracted from
  susceptibility data for Sr$_2$Cu$_{1-x}$Pd$_x$O$_{3}$
  \cite{Kojima}. The nominal Pd concentration is $x=0.5\%$ (black
  circles), $x=1\%$ (red squares), and $x=3\%$ (blue diamonds). The
  solid lines denote the theoretical result (\ref{Curie_constant}) for
  $p=1.2\%$, $p=1.4\%$, and $p=2.4\%$ (from bottom to top).}
\label{Fig_Curie_exp}
\end{center}
\end{figure}

Here we want to go one step further and compare the effective Curie
constant extracted from the experimental data
\begin{equation}
\label{Curie_exp}
C^{\mbox{\tiny exp.}} = \frac{T}{p}\l[\chi^{\mbox{\tiny exp.}}-\chi' - (1-p)\chi_{\bulk}\r]
\end{equation}
with the theoretical prediction. We will use the same values for the
concentration of chain breaks $p$ and the constant contribution to the
susceptibility $\chi'$ as in Table \ref{tab1}. For $\chi_\bulk$ we employ the
field theoretical result (\ref{bulk_thermo}). The result is shown in
Fig.~\ref{Fig_Curie_exp}.
Although the agreement is far from perfect we see that the Curie
constant extracted from experiment does indeed show a non-trivial
temperature dependence.  In particular, there is a finite temperature
minimum although at higher temperatures than predicted
theoretically. The maximum at low temperatures and subsequent drop of
the Curie constant, on the other hand, is clearly caused by the
interchain couplings which can no longer be neglected in this
temperature regime. Given that the procedure to extract the Curie
constant is quite sensitive to the values of $J$, $p$, and $\chi'$
used the agreement is quite remarkable supporting that the Curie-like
contribution is mainly caused by an effective concentration of chain
breaks $p$ and that possible additional paramagnetic Curie
contributions are relatively small.

For a more detailed comparison with theory, it would be desirable to
repeat the experiment with annealed samples, carefully estimating any
residual true paramagnetic Curie terms in the pure system, so that the
nominal Pd concentration directly corresponds to the concentration of
chain breaks created. 

\section{Conclusions}
\label{conclusions}
We have studied the effect of chain breaks (non-magnetic impurities)
on the thermodynamic properties of spin-$1/2$ chains by field theory
methods. Using first order perturbation theory in the leading
irrelevant operator (Umklapp scattering) we have derived a
parameter-free result for the thermodynamic properties of a finite
length $XXZ$-chain with open boundaries beyond the scaling limit. We
have shown that this result reduces to the previously known
expressions for boundary contributions in the limit of infinite chain
length. To obtain results for the isotropic case, we expanded the
results for the anisotropic model in terms of $1-1/K$ and re-expressed
this expansion in terms of a small running coupling constant $g$
fulfilling a set of known RG equations. For boundary contributions
like the boundary susceptibility $\chi_B$ this leads to multiplicative
logarithmic corrections. Whereas results in first order in $g$ can be
easily obtained by just replacing the Umklapp scattering amplitude
$\lambda\to g/4$, the consistent calculation of the correction of
order $g^n$ generally requires $n$-th order perturbation in the
Umklapp scattering. For the boundary susceptibility we presented the
result up to order $g^2$ using second order perturbation theory in
$\lambda$.

For the susceptibility of a finite length isotropic Heisenberg chain we were
able to derive a simple formula by studying the ground-state limit $LT/v\to
0$. In this limit the first order correction in the Umklapp scattering can be
absorbed into the scaling limit result. Surprisingly, we find by comparing
with QMC data that the simple parameter-free formula for $\chi(L,T)$ obtained
this way does work very well over a much larger range of temperatures and
lengths than anticipated. This simple formula is useful to analyse
susceptibility data for finite length Heisenberg chains.

We then discussed the average susceptibility $\chi_p$ for impurity
concentration $p$ assuming a Poisson distribution. Using the results
for $\chi(L,T)$ in the limiting cases $LT/v\to 0$ and $LT/v\to\infty$
we derived an effective fitting formula for $\chi_p$ which reproduces
QMC results with high accuracy. This formula can be used to fit
experimental data thus allowing to determine the (effective) impurity
concentration.

Taking into consideration that any real material will have interchain
$\tilde{J}$ as well as longer ranged intrachain couplings $J'$ which
can bridge over a chain break we discussed the limitations of the
theoretical analysis for $\chi_p$ presented here. Our main result is
that such couplings can be ignored provided that $T\gg
\mbox{Min}(\tilde{J},p^2J')$. Note, that $J'$ is suppressed by a
factor $p^2$ so that even relatively large couplings between the
segments can be ignored provided that the impurity concentration is
small. In most materials it will therefore be the loss of
one-dimensionality due to $\tilde{J}$ which marks the point where our
theory no longer applies.

For Sr$_2$CuO$_3$ this scale is about $5$ K whereas $J\sim 2000$ K
making it an ideal candidate to study the thermodynamics of impurities
in a Heisenberg chain experimentally over a large temperature
range. In this work we concentrated on analysing susceptibility data
by Kojima {\it et al.} \cite{Kojima} for
Sr$_2$Cu$_{1-x}$Pd$_x$O$_{3}$. We extracted the effective Curie
constant and showed that it agrees qualitatively with the theoretical
prediction. The impurity concentration, however, seems to be different
from the nominal Pd concentration. A plausible explanation is the
presence of excess oxygen leading to additional chain breaks, a
scenario already discussed in \cite{AmiCrawford,MotoyamaEisaki}. It
might be possible to avoid these additional chain breaks by annealing
which seems to make a more detailed comparison with theory feasible in
the future.

Our results might also shed light on some aspects of the high-$T_c$
cuprates physics. In YBa$_2$Cu$_3$O$_{6+\delta}$ we have
quasi-one-dimensional CuO chains in addition to the CuO$_2$ planes
common to all high-$T_c$ cuprates. By changing the oxygen content
$\delta$ the charge concentration in the planes is changed. At the
same time, this leads to the removal of spins from the chain. If we
assume a Zhang-Rice singlet type state with immobile holes this again
leads to chain breaks in the same way as in
Sr$_2$CuO$_{3+\delta}$. Indeed, Friedel-type spin-density oscillations
have been observed in YBa$_2$Cu$_3$O$_{6.5}$ by NMR
\cite{YamaniStatt}. Such oscillations are expected if spin chain
segments with open boundaries are present \cite{EggertAffleck95}. A
thorough analysis of these data and possible new experiments on
Sr$_2$Cu$_{1-x}$Pd$_x$O$_{3}$ or similar compounds might give us a
better understanding of oxygen doping and the formation of Zhang-Rice
singlet states in the cuprates.

\section*{Acknowledgments}
We would like to thank M.~Bortz, F.H.L.~Essler, N.~Kawakami, and I. Schneider
for valuable discussions.  We are also grateful to A.~Furusaki for sharing his
insight about related issues and to F.~Anfuso for making his Monte Carlo code
available to us for part of the simulations. This work was partly supported by
a Grant-in-Aid from the Ministry of Education, Science, Sports and Culture,
Japan (SF), the Swedish Research Council (SE), the Transregio 49 funded by the
DFG (SE), NSERC (IA), and CIfAR (IA). JS thanks the University of New South
Wales, Sydney, Australia for their hospitality and the Gordon Godfrey fund for
financial support. NL acknowledges computer facilities provided by the
WestGrid network, funded in part by the Canada Foundation for Innovation.

\appendix
\section{Second order perturbation theory}
\label{app_A}
The free energy in second order in the Umklapp scattering is given by
\begin{eqnarray}
\label{A1}
\fl f_2 &=& -\frac{\lambda^2}{2}\frac{T}{L}\l\{ \int_0^L \int_0^\beta d^2x\, d^2\tau \r. \\
\fl &\times& \l\langle\cos\l(\sqrt{8\pi K}\phi(x_1,\tau_1)+\frac{2Khx_1}{v}\r)\cos\l(\sqrt{8\pi K}\phi(x_2,\tau_2)+\frac{2Khx_2}{v}\r)\r\rangle  \nonumber \\
\fl &-& \l.\l[\int_0^L\int_0^\beta \l\langle \cos\l(\sqrt{8\pi K}\phi(x_1,\tau_1)+\frac{2Khx_1}{v}\r)\r\rangle \r]^2 \r\} \; . \nonumber
\end{eqnarray}
We want to evaluate this contribution here only in the thermodynamic
limit $L\to\infty$ where it splits into a bulk and a boundary
contribution. In this limit the two-point correlation function for OBC
at zero temperature is given by
\begin{eqnarray}
\label{A2}
\fl \langle \phi(x_1,\tau_1) \phi(x_2,\tau_2)\rangle && \\
 &\!\!\!\!\!\!\!\!\!\!\!\!\!\!\!\!\!\!\!\!\!\!\!\!\!\! =&\!\!\!\!\!\!\!\!\!\!\!\!\!
-\frac{1}{4\pi}\l(\ln\l[(x_1-x_2)^2+v^2(\tau_1-\tau_2)^2\r]-\ln\l[(x_1+x_2)^2+v^2(\tau_1-\tau_2)^2\r]\r)
\, .\nonumber
\end{eqnarray}
The exponentials of the bosonic field $\phi$ can then by obtained by
using the cumulant theorem
\begin{equation}
\label{A3}
\fl\l\langle\exp\l(\pm\im\sqrt{8\pi K}\phi(x,t)\r)\r\rangle = \exp\l(-4\pi K\langle\phi(x,t)\phi(x,t)\rangle\r) 
\end{equation}
and
\begin{eqnarray}
\label{A4}
\fl
\l\langle\exp\l(\im\sqrt{8\pi K}\phi(x_1,\tau_1)\r)\exp\l(\pm\im\sqrt{8\pi K}\phi(x_2,\tau_2)\r)\r\rangle \\
\fl \qquad =\exp\l(-4\pi K[\langle\phi^2(x_1,\tau_1)\rangle +
\langle\phi^2(x_2,\tau_2)\rangle \pm 2
\langle\phi(x_1,\tau_1)\phi(x_2,\tau_2)\rangle ]\r) \; . \nn
\end{eqnarray}
Using the fact that $\langle\sin(\sqrt{8\pi K}\phi)\rangle\equiv 0$, the
susceptibility is given by
\begin{eqnarray}
\label{A5}
\fl \chi_2 &=& -2\lambda^2\frac{T}{L}\l(\frac{K}{v}\r)^2\l\{\int
\int d^2x\, d^2\tau \r. \nn \\
\fl &\times& \l[(x_1^2+x_2^2)
\l\langle\cos\l(\sqrt{8\pi K}\phi(x_1,\tau_1)\r)\cos\l(\sqrt{8\pi K}\phi(x_2,\tau_2)\r)\r\rangle\r. \nn
\\ 
\fl &-& \l. 2x_1x_2
\l\langle\sin\l(\sqrt{8\pi K}\phi(x_1,\tau_1)\r)\sin\l(\sqrt{8\pi K}\phi(x_2,\tau_2)\r)\r\rangle\r]
\\
\fl &-& \l.\frac{2}{T^2}\l[\int dx
\l\langle\cos\l(\sqrt{8\pi K}\phi(x)\r)\r\rangle\r]\l[\int dx\,
x^2\l\langle\cos\l(\sqrt{8\pi K}\phi(x)\r)\r\rangle\r]\r\} \; .\nn
\end{eqnarray}
Using (\ref{A3}, \ref{A4}) the correlation functions can be
evaluated. The usual conformal mapping then allows to obtain the
second order correction at finite temperature
\begin{eqnarray}
\label{A6}
\fl \chi_2 &=& -\frac{\lambda^2}{L}\frac{K^2}{v^3}\l(\frac{\pi
  T}{v}\r)^{4K-5}\int_0^{\pi TL/v}d^2x\int_0^\pi d\tau \\
\fl &\times& \l\{ (x_1-x_2)^2
\l[\frac{\sinh(x_1+x_2+\im\tau)\sinh(x_1+x_2-\im\tau)}{\sinh(2x_1)\sinh(2x_2)\sinh(x_1-x_2+\im\tau)\sinh(x_1-x_2-\im\tau)}\r]^{2K}\r.
\nn \\
\fl &+& (x_1+x_2)^2
\l[\frac{\sinh(x_1-x_2+\im\tau)\sinh(x_1-x_2-\im\tau)}{\sinh(2x_1)\sinh(2x_2)\sinh(x_1+x_2+\im\tau)\sinh(x_1+x_2-\im\tau)}\r]^{2K}
\nn \\
\fl &-& \l. 2(x_1+x_2)^2
\l[\frac{1}{\sinh(2x_1)\sinh(2x_2)}\r]^{2K}\r\} \; .\nn
\end{eqnarray} 
If we are far from the boundary, $x_1,x_2\gg 1$, then only the first
term contributes and the integral gives a contribution to the bulk susceptibility
\begin{eqnarray}
\label{A7}
 \chi_2^{\mbox{\tiny bulk}} &=& -\frac{\lambda^2}{L}\frac{K^2}{v^3}\l(\frac{\pi
  T}{v}\r)^{4K-5}\int_0^{\pi TL/v}d^2x\int_0^\pi d\tau \nn \\
 &\times&  (x_1-x_2)^2
\l[\frac{1}{\sinh(x_1-x_2+\im\tau)\sinh(x_1-x_2-\im\tau)}\r]^{2K}
\\
&\to& -\lambda^2\frac{K^2}{v^3}\l(\frac{\pi
  T}{v}\r)^{4K-4}\int_0^\infty dx \int_0^\pi d\tau
\frac{x^2}{\l[\frac{1}{2}(\cosh{2x}-\cos{2t})\r]^{2K}} \; . \nn
\end{eqnarray} 
This integral can be evaluated analytically yielding the known result
\cite{Lukyanov,SirkerBortzJSTAT} 
\begin{eqnarray}
\label{A8}
\fl \chi_2^{\mbox{\tiny bulk}} = \frac{\lambda^2}{32\pi
  v^3}\Gamma^2(1/2-K)\Gamma^2(1+K)\sin(2\pi
K)\l[\Psi'(1-K)-\Psi'(K)\r]\l(\frac{\pi T}{v}\r)^{4K-4} \!\!\!\!\! .
\end{eqnarray} 
To obtain the boundary contribution $\chi_2^B$ we have to subtract
(\ref{A7}) from (\ref{A6}). As the kernel will then go to zero away
from the boundary, we can shift the upper bound in the integration to
infinity leading to
\begin{eqnarray}
\label{A9}
\fl \chi_2^B  &=& \chi_2 - \chi_2^{\mbox{\tiny bulk}} = -\frac{\lambda^2}{L}\frac{K^2}{v^3}\l(\frac{\pi
  T}{v}\r)^{4K-5}\int_0^{\infty}d^2x\int_0^\pi d\tau \\
\fl &\times&\l[\sinh 2x\sinh 2y\l(\cosh 2(x+y)-\cos 2t\r)\l(\cosh
2(x-y)-\cos 2t\r)\r]^{-2K} \nn \\
\fl &\times& \l\{ (x_1^2+x_2^2)\l[\l[ \l(\cosh 2(x+y)-\cos 2t\r)^{2K}
-\l(\cosh 2(x-y)-\cos 2t\r)^{2K}\r]^2 \r.\r. \nn \\
\fl &-& \l.\l[2\sinh 2x\sinh 2y\l(\cosh
2(x+y)-\cos 2t\r)\r]^{2K}\r] \nn \\
\fl &-& 2x_1x_2 \l[\l(\cosh 2(x+y)-\cos 2t\r)^{4K}
-\l(\cosh 2(x-y)-\cos 2t\r)^{4K}\r. \nn \\
\fl &-&\l.\l. \l[2\sinh 2x\sinh 2y\l(\cosh
2(x+y)-\cos 2t\r)\r]^{2K}\r]\r\} \nn 
\end{eqnarray}
The integral is convergent for $K<3/2$ ($\Delta>1/2$) and can be
evaluated numerically.
%% Delta      2*Integral
%% 0.6        24.67
%% 0.7        20.85
%% 0.8        8.16
%% 0.9        1.68
%% 1.0        -1.66
Note, that this is the result for a semi-infinite chain, i.e., this
result has to be multiplied by a factor 2 for an open chain segment.

\section{RG-improved coupling constants in the isotropic case}
\label{app_B}
The coupling constants in (\ref{int_iso}) fulfil the following set of
RG equations \cite{Lukyanov}
\begin{equation}
\label{B1}
\frac{dg_\parallel}{d\ln r} =-\frac{2g_\perp^2}{2-g_\parallel}\;
,\qquad \frac{dg_\perp}{d\ln r} =-\frac{2g_\parallel
  g_\perp}{2-g_\parallel} \; .
\end{equation}
Here $r$ denotes the appropriate RG scale ($T$ or $L$ in the cases
considered here). Following Lukyanov \cite{Lukyanov} the solution to
this set of equations can be parametrised as
\begin{equation}
\label{B2}
g_\parallel = 2(1-1/K)\frac{1+q}{1-q}\; , \qquad g_\perp = -4(1-1/K)\frac{\sqrt{q}}{1-q}
\end{equation}
where
\begin{equation}
\label{B3}
q(1-q)^{2K-2}=\l(\frac{r}{r_0}\r)^{4-4K} \; .
\end{equation}
Here $r_0$ is a constant which can be chosen freely.  

This is the setup which quite generally allows to obtain results for
the isotropic model by expanding the results for the anisotropic model
in a power series in $(1-1/K)$. From the scaling part (\ref{scalechi})
we obtain for the bulk susceptibility
\begin{equation}
\label{B4}
\chi_{\mbox{\tiny bulk}}^{(0)} =\frac{K}{2\pi v} =\frac{K}{\pi^2}
=\frac{1}{\pi^2}\sum_{n=0}^\infty (1-1/K)^n
\to\frac{1}{\pi^2}\sum_{n=0}^\infty \l(\frac{g_\parallel}{2}\r)^n \; .
\end{equation}
Note, that the power series in $g_\parallel$ produces terms higher
order in $q$ apart from the $q^0$-term required. In particular,
$g_\parallel/2 = (1-1/K)(1+2q+\mathcal{O}(q^2)$ with $(1-1/K)q\sim
(1-1/K) T^{4K-4}$. Such a term is indeed present when expanding the
second order contribution in $\lambda$, Eq.~(\ref{A8}), in powers of
$(1-1/K)$ and fixes the scale $r_0$ in (\ref{B3}). In other words, the
scale in (\ref{B3}) is uniquely fixed because we have two equations
for the two unknowns: The pre-factor of $g_\parallel$ is fixed through
(\ref{B4}) and the expansion of (\ref{A8}) fixes the scale to
\begin{equation}
\label{B5}
q(1-q)^{2K-2}
=\l[\frac{T\e^{-1/4}\sqrt{\pi}\,\Gamma\l(1+\frac{1}{2K-2}\r)}{v\,\Gamma\l(1+\frac{K}{2K-2}\r)}\r]^{4K-4}\frac{\Gamma^2(K)}{\Gamma^2(2-K)}
\; .
\end{equation}
Comparing with the expression for the coupling constant $\lambda$,
Eq.~(\ref{B108}), we see that we can write
\begin{equation}
\label{B6}
\lambda = (K-1)\sqrt{q}(1-q)^{K-1}\e^{(K-1)/2}\l[\frac{v}{2\pi T}\r]^{2K-2}
\end{equation}
which in lowest order in $(1-1/K)$ and $q$ reduces to 
\begin{equation}
\label{B7}
\lambda = - \frac{g_\perp}{4} \; .
\end{equation}

Next, we consider the first order result for the boundary
susceptibility (\ref{chi1}). Using (\ref{B6}) we find
\begin{equation}
\label{B8}
\fl \delta\chi_1 = K^2(1-K)\e^{(K-1)/2}B(K,1-2K)\l[\pi^2-2\psi'(K)\r]\frac{\sqrt{q}(1-q)^{K-1}}{\pi^2
  T} \; .
\end{equation}
Expanding this expression in $(1-1/K)$ and $q$ yields
\begin{equation}
\label{B8b}
\fl T\,\delta\chi_1 =
\frac{(1-1/K)}{3}\sqrt{q}+(1-1/K)^2\l(\frac{1}{2}-\frac{\psi''}{\pi^2}\r)\sqrt{q}
-\frac{(1-1/K)^2}{3}q^{3/2} +\cdots \; .
\end{equation}
For the first two terms we can write
\begin{equation}
\label{B9}
 T\,\delta\chi_1 = -\frac{g_\perp}{12}-\frac{g_\parallel
  g_\perp}{8}\l(\frac{1}{2}-\frac{\psi''}{\pi^2}\r) \; ,
\end{equation}
however, this does not reproduce the third term in (\ref{B9})
correctly. To obtain agreement also in order $(1-1/K)^2q^{3/2}$ the
scale in (\ref{B5}) has to be changed. This leads to minimal
improvements only, as this term is not only second order in $(1-1/K)$
but also third order in $\sqrt{q}$. We therefore keep the scale as
determined from the expansion of the bulk susceptibility,
Eq.~(\ref{B5}), which is convenient to describe $\chi(L,T)$ at the
isotropic point. More important than the third term in (\ref{B9}) is
actually the second order contribution (\ref{A9}) derived in the
previous section. This contribution can be expressed as
\begin{equation}
\label{B10}
 \chi_2^B =
\frac{4}{\pi^3T}K^2(K-1)^2\e^{K-1}q(1-q)^{2K-2}(-2I) \; , 
\end{equation} 
where $I$ denotes the integral in (\ref{A9}) and the factor $2$ has
been added to account for the two boundaries. To lowest order in
$(1-1/K)$ we therefore obtain
\begin{equation}
\label{B11}
 \chi_2^B =-\frac{8}{\pi^3T} (1-1/K)^2q \, I(K=1) \to -
 \frac{g_\perp^2}{2\pi^3 T} \, I(K=1)
\end{equation} 
and a numerical evaluation yields $I(K=1)\approx -0.83$. Finally, with
$g_\parallel\to g$ and $g_\perp \to -g$ Eqs.~(\ref{B9}) and (\ref{B11})
yield (\ref{chiB_iso3}).

$g$ as a function of temperature $T$ can be found as follows: Using
the first equation of (\ref{B2}) we can write
$q=[g-2(1-1/K)]/[g+2(1-1/K)]$. Now we can put this into
$q(1-q)^{2K-2}$ and expand in $(1-1/K)$ leading to 
\begin{equation}
\label{B12}
\fl q(1-q)^{2K-2} \approx
1+\l[-\frac{4}{g}+2+2\ln\l(\frac{4}{g}\r)+2\ln(1-1/K)\r](1-1/K) \; .
\end{equation} 
On the other hand, we can expand (\ref{B5}) leading to 
\begin{eqnarray}
\label{B13}
\fl &&
\l[\frac{T\e^{-1/4}\sqrt{\pi}\,\Gamma\l(1+\frac{1}{2K-2}\r)}{v\,\Gamma\l(1+\frac{K}{2K-2}\r)}\r]^{4K-4}\frac{\Gamma^2(K)}{\Gamma^2(2-K)}
\nn \\
\fl &\approx& 1+\l[1+2\ln 2+2\ln(1-1/K)-4\gamma-2\ln\pi+4\ln(2T/J)\r] \; .
\end{eqnarray} 
Comparing (\ref{B12}) and (\ref{B13}) finally leads to (\ref{coup_iso}).

%% \bibliography{Literatur}

\begin{thebibliography}{10}

\bibitem{Bethe}
Bethe H 1931 {\em Z. Phys.\/} {\bf 71} 205

\bibitem{BonnerFisher}
Bonner J and Fisher M 1964 {\em Phys. Rev.\/} {\bf 135} A640

\bibitem{Affleck98}
Affleck I 1998 {\em J. Phys. A\/} {\bf 31} 4573

\bibitem{Lukyanov}
Lukyanov S 1998 {\em Nucl. Phys. B\/} {\bf 522} 533

\bibitem{LukyanovTerras}
Lukyanov S and Terras V 2003 {\em Nucl. Phys. B\/} {\bf 654} 323

\bibitem{egg94}
Eggert S, Affleck I and Takahashi M 1994 {\em Phys. Rev. Lett.\/} {\bf 73} 332

\bibitem{EggertAffleck92}
Eggert S and Affleck I 1992 {\em Phys. Rev. B\/} {\bf 46} 10866

\bibitem{AsakawaMatsuda}
Asakawa H, Matsuda M, Minami K, Yamazaki H and Katsumata K 1998 {\em Phys. Rev.
  B\/} {\bf 57} 8285

\bibitem{WesselHaas}
Wessel S and Haas S 2000 {\em Phys. Rev. B\/} {\bf 61} 15262

\bibitem{FujimotoEggert}
Fujimoto S and Eggert S 2004 {\em Phys. Rev. Lett.\/} {\bf 92} 037206

\bibitem{FurusakiHikihara}
Furusaki A and Hikihara T 2004 {\em Phys. Rev. B\/} {\bf 69} 094429

\bibitem{EggertAffleckHorton}
Eggert S, Affleck I and Horton M 2003 {\em Phys. Rev. Lett.\/} {\bf 90} 89702

\bibitem{BrunelBocquet}
Brunel V, Bocquet M and Jolicoeur T 1999 {\em Phys. Rev. Lett.\/} {\bf 83} 2821

\bibitem{AffleckQin}
Affleck I and Qin S 1999 {\em J. Phys. A\/} {\bf 32} 7815

\bibitem{EggertRommer}
Eggert S and Rommer S 1998 {\em Phys. Rev. Lett.\/} {\bf 81} 1690

\bibitem{RommerEggert}
Rommer S and Eggert S 1999 {\em Phys. Rev. B\/} {\bf 59} 6301

\bibitem{RommerEggert2}
Rommer S and Eggert S 2000 {\em Phys. Rev. B\/} {\bf 62} 4370

\bibitem{EggertAffleck95}
Eggert S and Affleck I 1995 {\em Phys. Rev. Lett.\/} {\bf 75} 934

\bibitem{FHLE}
Essler F~H~L 1996 {\em J. Phys. A\/} {\bf 29} 6183

\bibitem{AsakawaSuzuki96a}
Asakawa H and Suzuki M 1996 {\em J. Phys. A.\/} {\bf 29} 225

\bibitem{AsakawaSuzuki96b}
Asakawa H and Suzuki M 1996 {\em J. Phys. A.\/} {\bf 29} 7811

\bibitem{frah97}
Frahm H and Zvyagin A~A 1997 {\em J. Phys. Cond. Mat.\/} {\bf 9} 9939

\bibitem{Fujimoto}
Fujimoto S 2000 {\em Phys. Rev. B\/} {\bf 63} 024406

\bibitem{BortzSirker}
Bortz M and Sirker J 2005 {\em J. Phys. A: Math. Gen.\/} {\bf 38} 5957

\bibitem{SirkerBortzJSTAT}
Sirker J and Bortz M 2006 {\em J. Stat. Mech.\/}  P01007

\bibitem{SirkerLaflorencie}
Sirker J, Laflorencie N, Fujimoto S, Eggert S and Affleck I 2007 {\em Phys.
  Rev. Lett.\/} {\bf 98} 137205

\bibitem{AmiCrawford}
Ami T, Crawford M~K, Harlow R~L, Wang Z~R, Johnston D, Huang Q and Erwin R 1995
  {\em Phys. Rev. B\/} {\bf 51} 5994

\bibitem{MotoyamaEisaki}
Motoyama N, Eisaki H and Uchida S 1996 {\em Phys. Rev. Lett.\/} {\bf 76} 3212

\bibitem{Kojima}
Kojima K and {\it et al} 2004 {\em Phys. Rev. B\/} {\bf 70} 094402

\bibitem{TakigawaMotoyama}
Takigawa M, Motoyama N, Eisaki H and Uchida S 1997 {\em Phys. Rev. B\/} {\bf
  55} 14129

\bibitem{ThurberHunt}
Thurber K~R, Hunt A~W, Imai T and Chou F~C 2001 {\em Phys. Rev. Lett.\/} {\bf
  87} 247202

\bibitem{Affleck_lesHouches}
Affleck I 1988 {\em Fields, Strings and Critical Phenomena\/} Les Houches,
  Session XLIX, Amsterdam p.~563 edited by E. Br\'ezin and J. Zinn-Justin

\bibitem{tak99}
Takahashi M 1999 {\em Thermodynamics of one-dimensional solvable problems\/}
  Cambridge University Press

\bibitem{Sandvik2002}
Syljuasen O~F and Sandvik A~W 2002 {\em Phys. Rev. E\/} {\bf 66} 046701

\bibitem{RosnerEschrig}
Rosner H, Eschrig H, Hayn R, Drechsler S~L and M\'alek J 1997 {\em Phys. Rev.
  B\/} {\bf 56} 3402

\bibitem{Fisher_IRFP}
Fisher D~S 1994 {\em Phys. Rev. B\/} {\bf 50} 3799

\bibitem{HillJohnston}
Hill J~M, Johnston D~C and Miller L~L 2002 {\em Phys. Rev. B\/} {\bf 65} 134428

\bibitem{YamaniStatt}
Yamani Z, Statt B~W, MacFarlane W~A, Liang R, Bonn D~A and Hardy W~N 2006 {\em
  Phys. Rev. B\/} {\bf 73} 212506

\end{thebibliography}

\end{document}